\documentclass[epj]{svjour}
\usepackage{graphics}
\usepackage{xcolor}
\usepackage{amsmath}
\usepackage{xspace}
\usepackage{slashed}
\usepackage{subcaption}
\usepackage{cite}
\RequirePackage{listings}
\usepackage{listings}
\usepackage{booktabs} %\midrule
\usepackage{enumitem} % roman numerals in enumerate
\usepackage[normalem]{ulem}
\usepackage{def}

\definecolor{codegreen}{rgb}{0,0.6,0}
\definecolor{codegray}{rgb}{0.5,0.5,0.5}
\definecolor{codepurple}{rgb}{0.58,0,0.82}
\definecolor{backcolor}{rgb}{0.95,0.95,0.92}
\definecolor{tabgrey}{rgb}{0.57,0.57,0.57}
\definecolor{myblack}{rgb}{0,0,0}

\lstdefinestyle{mystyle}{
    backgroundcolor=\color{backcolor},   
    commentstyle=\color{codegreen},
    keywordstyle=\color{myblack},
    numberstyle=\tiny\color{codegray},
    stringstyle=\color{codepurple},
    basicstyle=\ttfamily\footnotesize,
    breakatwhitespace=false,         
    breaklines=true,                 
    captionpos=b,                    
    keepspaces=true,                 
    numbers=left,                    
    numbersep=5pt,                  
    showspaces=false,                
    showstringspaces=false,
    showtabs=false,                  
    tabsize=4
}
\definecolor{darkmagenta}{rgb}{0.55, 0.0, 0.55}
\begin{document}
%
% \title{Implementing Berends-Giele-Like Recursions and a $1/\Nc$ Expansion in \mg}
\title{Improving Colour Computations in \mg{} and Exploring a $1/\Nc$ Expansion}
% \subtitle{Do you have a subtitle?\\ If so, write it here}
\author{Andrew Lifson\inst{1,2} \and Olivier Mattelaer\inst{2}% etc
% \thanks{\emph{Present address:} Insert the address here if needed}%
}                     % Do not remove
%
% \offprints{}          % Insert a name or remove this line
%
\institute{Department of Astronomy and Theoretical Physics, Lund University, Sölvegatan 14A, 223 62 Lund, Sweden \and 
Centre for Cosmology, Particle Physics and Phenomenology (CP3), Université catholique de
Louvain, Chemin du Cyclotron 2, B-1348 Louvain-la-Neuve, Belgium}
\date{Received: date / Revised version: date}
\abstract{
In this paper, we present an extension of \mg{}
which is able to evaluate tree-level QCD matrix-elements up to $2\to 6$
(one more particle than before). 
To achieve this, 
we implemented Berends-Giele-like recursion,
and re-implemented the way colour is computed such that we can now expand the colour matrix in powers of 
$1/\Nc$ and truncate this expansion to a chosen order. 
% 
% allowing to choose which terms of the colour-matrix to include and allowing us to study the $1/\Nc$
% % This implementation allows to choose which terms of the colour-matrix to include and allows us to study the $1/\Nc$
% expansion of the colour-matrix. 
For high multiplicity samples, 
even without truncating the colour matrix,
the new implementation offers a speed gain compared to the previous \mg{} code.
 % end of PACS codes
} %end of abstract

\authorrunning{A. Lifson \& O. Mattelaer}
\titlerunning{Improving colour computations in \mgs{}}

\maketitle
%
% LU-TP 22??, MCNET-22-??,CP3-22-42
\tableofcontents

\section{Introduction}
\label{sec:introduction}

% \begin{itemize}
%     \item Automation of MEs important to theory/predictions
%     \item There exists many programs to do this, list programs
%     \item Perhaps something about the computer budget for next 10 years is smaller than expected usage, so need to find ways to improve speed of calculations
%     \item Mention that we will get more high-multiplicity final states
%     \item Mention that Feynman diagrams are inherently slow, i.e.\ factorial growth issue
%     \item Mention that colour is a major bottleneck \cite{Mattelaer:2021xdr} and directions taken to address this
%     \item Berends-Giele known to be faster than standard Feynman diagrams.
%     \item Perhaps mention why BG not already in \mg{} and why this issue no longer exists?
%     \item Introduce rest of paper.
% \end{itemize}

Accurate and efficient calculation of the hard matrix element is at the core of most predictions in high-energy physics,
with many tools currently available to automate these calculations 
\cite{Kanaki:2000ey,Moretti:2001zz,Krauss:2001iv,Mangano:2002ea,Gleisberg:2008fv,Berger:2008sj,Alwall:2014hca,Kilian:2007gr,Belyaev:2012qa,Hahn:2000kx,Cafarella:2007pc,Bellm:2015jjp,Cascioli:2011va,Bevilacqua:2011xh,Badger:2012pg,Cullen:2014yla,Actis:2016mpe,Denner:2017wsf,Honeywell:2018fcl,Buccioni:2019sur,Hirschi:2011pa,Denner:2016kdg,Hirschi:2016mdz,Ossola:2007ax,vanHameren:2010cp,Ellis:2007qk}. 
However, due to a much faster growth in the number of events    required for the high-luminosity LHC compared to the growth of the LHC cpu-hour budget,
the efficiency of such programs needs to be improved by at least 20 percent and ideally by a factor of two
\cite{HEPSoftwareFoundation:2020daq,Collaboration:2802918}

% However, due to the vast number of events we need to generate for the high-luminosity LHC era, the efficiency of such programs needs to be improved by at least 20 percent and ideally by a factor of two, to meet the LHC need within the budget of cpu-hour available \cite{HEPSoftwareFoundation:2020daq}.

%  and the relatively small LHC cpu-hour budget
% \AL{Don't like this sentence, it is too disconnected from previous paragraph.}
% Many interesting processes in the high energy and high luminosity LHC will involve multiple jets.
At the high-luminosity LHC, we expect to see and generate many processes with multiple well-separated jets.
There are two challenges to calculating the hard matrix element for these types of processes,
even at tree level. 
The first is to quickly calculate the Lorentz part of the amplitude (often called the kinematic part), 
typically calculated by summing Feynman diagrams which grow roughly factorially with the number of external particles;
while the second is to calculate the colour algebra, which typically grows like a factorial squared with the number of external particles.
Indeed, as the multiplicity increases, the colour takes up a larger percentage of a  \mg{} (MG5\-aMC) calculation,
with the colour taking up about 60\% of the time required to calculate the cross section of $gg\rightarrow t\bar{t} g g g$ \cite{Mattelaer:2021xdr}.

There have been several attempts to speed up the Feynman diagrams used to calculate the kinematics \cite{Mattelaer:2021xdr,Lifson:2020pai,Alnefjord:2020xqr,Lifson:2022ijv,Maitre:2021uaa,Ballestrero:1994jn}.
However, an alternative method to speed up the kinematics, is to use recursions such as the off-shell Berends-Giele (BG) recursion instead of Feynman diagrams \cite{Berends:1987me}.
These recursions sum up multiple Feynman diagrams into a single term, thus decreasing the required amount of computation,
and have already been implemented in e.g.\ \cite{Gleisberg:2008fv,Cafarella:2007pc}.
Other recursive methods can include other off-shell recursions such as \cite{Kosower:1989xy,Schwinn:2005pi}, 
% \commentAL{Scalar vertices and Bern paper}
or on-shell recursion relations such as \cite{Britto:2004ap,Britto:2005fq,Cachazo:2004kj}, 
% \commentAL{BCFW, CSW, others?},
though past studies have shown that BG recursions are typically quicker \cite{Dinsdale:2006sq,Badger:2012uz,Gleisberg:2008ft}.

For the colour, research has mainly focused on two directions. 
The first is to diagonalise the colour matrix, thus severely reducing the number of elements in the colour matrix.
This is mostly realised in the multiplet basis \cite{Keppeler:2012ih,Sjodahl:2015qoa,Sjodahl:2018cca}.
The second approach is to use the large-$\Nc$ limit \cite{tHooft:1973alw},
% \commentAL{Any other citations for large Nc? I can only think that it's used in Parton showers and as a first step in higher-order calculations etc. so we could mention that it's widely used, then go back to this in the conclusions}
and expand the colour matrix in a power series in $1/\Nc$.
For the most relevant processes, each order of the expansion is separated by two powers of $1/\Nc$,
making the expansion about as accurate as the expansion in $\alpha_s$.

In this paper, we implement both BG recursion and a colour expansion in \mgs{}
for tree-level Standard Model processes.
In \secref{sec:background}, we summarise colour ordering and the $1/\Nc$ 
expansion, as well as BG recursions. 
Next, in \secref{sec:implementation},
we describe and profile their implementations in the \mgs{} event generator.
We show our results for pure QCD processes in \secref{sec:validation and results},
showing the accuracy and speed of the colour expansion. 
% \AL{and the speed of the new program relative to the old one at each order of the colour expansion}. 
We conclude in \secref{sec:conclusion}.
A small user manual is described in \appref{sec:manual}.
In \appref{sec:EW results},
we briefly show our results for some additional processes including those with an electroweak boson. 
We study the relative importance of different subprocesses in a typical QCD cross section in \appref{mlm}.
Finally, in \appref{sec:modLC multiquark},
we describe a proposed modified definition of the colour expansion in multiquark amplitudes.

\section{Background Theory}
\label{sec:background}
In this section, we describe the two main ideas we implemented in this paper,
colour ordering in the fundamental basis and its expansion in powers of $1/\Nc$,
and the use of Berends-Giele recursion to calculate colour-ordered kinematic amplitudes.

\subsection{Colour Ordering and the $1/\Nc$ Expansion}
\label{sec:col ordering}
% \begin{itemize}
%     \item Comment on general factorisation of colour and kinematics
%     \item Show more concretely the factorisation in the fundamental basis
%     \item Subsection for all-gluon amplitudes since we don't take the strict fundamental 
%     $1/\Nc$ expansion there?
%     \item Subsection for multi-quark amplitudes since we don't take the strict fundamental 
%     $1/\Nc$ expansion there? 
%     (Also mention here that expansion is not as obvious since the next order terms only one power of 
%     $\Nc$ smaller than current order?)
% \end{itemize}
% \AL{remove the paragraph below? Seems to be unnecessary}
% Here, we describe colour ordering and the $1/\Nc$ expansion used throughout the paper.
% This ordering is done using the fundamental basis with a slight modification for the leading-colour all-gluon amplitude.
% First, we describe colour ordering in general and give the expressions for the colour-ordered amplitudes used,
% and then we describe how we expand the colour matrix in powers of $1/\Nc$.

\subsubsection{Colour Ordering in the Fundamental Basis}
\label{sec:col ordering fundamental}

A trick which is often used in QCD calculations is to factorise the colour part of an amplitude from the kinematics
\cite{Mangano:1987xk,DelDuca:1999iql,DelDuca:1999rs,Maltoni:2002mq}
\begin{equation}
\mathcal{M}(1,\dots,n) = \sum_{\sigma}F_\si(\text{su}(\Nc))M_\si(p_1,h_1;\dots;p_n,h_n)~,
\label{eq:colour ordering gen}
\end{equation}
% \OM{do not use f,t, but SU(N)}
where, for e.g.\ the fundamental (also called the trace) or colour-flow bases, 
$\sigma$ is a given permutation of colour orderings,
$F_\si$ is a function of the gauge algebra su($\Nc$),
and $M_\si$ is the kinematic (colour ordered) amplitude, 
which is a function of the momenta and helicities of the particles.
Depending on the basis in colour space, 
there may be different forms of $F_\si$ and $M_\si$, and different sets of permutations $\si$.

The squared matrix-element is then given by, 
\begin{equation}
\vert\mathcal{M}(1,\dots,n)\vert^2 = \sum_{\si,\si'}M_{\si}
\underbrace{F_{\si}F^*_{\si'}}_{C_{\si\si'}} M^*_{\si'}~,
\label{eq:colour sum}
\end{equation}
where we have dropped all functional dependence on the right hand side;
$\si,\si'$ are two sets of colour-ordering permutations; 
and the product $F_{\si}F^*_{\si'} \equiv C_{\si\si'}$ is called the colour matrix, which in e.g.\ the fundamental or colour flow bases is a square matrix with size growing factorially with the particle multiplicity.
The colour matrix typically contains polynomials in $\Nc$, the number of colours, and is calculated using the following colour-algebra relations:
\begin{align}
\Tr(t^a) &= 0~,
&
\Tr(t^at^b) &= \TR\de^{ab} ~, \nonumber \\
if^{abc} &= \frac{1}{\TR} \Tr(t^a[t^b,t^c])~,
& 
if^{abc}t^c &= [t^a,t^b]~, \nonumber \\
\de_{ii} &= \Nc~,
&
\de^{aa} &= \Nc^2-1~, \nonumber \\
t^a_{ij}t^a_{kl} &= \TR\left( \de_{il}\de_{jk} - \frac{1}{\Nc}\de_{ij}\de_{kl} \right)~.
\label{eq:colour algebra}
\end{align}
Here, $i,j,k,l = 1,2,\dots,\Nc$ are (anti)fundamental indices,
$a,b,c = 1,2,\dots,\Nc^2-1$ are adjoint indices,
all repeated indices are summed,
$t^a_{ij}$ is a generator of $\mathrm{su}(\Nc)$ in the fundamental representation,
$f^{abc}$ the structure constants (or equivalently the generators of
$\mathrm{su}(\Nc)$ in the adjoint representation),
and $\TR$ is a normalisation factor, in \mgs{} set to $1/2$ 
(though in the literature it is often set to one).
% , e.g.\ in \cite{}
% \commentAL{Mangano and Parke? something by Dixon? focus on lecture notes?}\OM{I would not put citation here}). 

In \mgs{}, the fundamental basis is used to calculate the colour matrix.
In this basis, all colour factors are written as strings of fundamental matrices $t_{ij}^a$.
For example, the all-gluon amplitude is written as 
\begin{equation}
\mathcal{M}(ng) = \sum_{P(2,\dots,n)}\Tr(t^1\dots t^n)M(1,\dots,n)~,
\label{eq:allGlue amplitude fundamental}
\end{equation}
where, $P(2,\dots,n)$ indicates the sum of all permutations of particles $2\dots n$
(particle $1$ is fixed to not double count, since the trace is cyclic).

This gives a colour matrix $C^{ng}_{\si\si'}$
\begin{align}
C^{ng}_{\si\si'} &= \Tr(t^{\si_1}\dots t^{\si_n})\Tr(t^{\si'_n}\dots t^{\si'_1})~,
\label{eq:col matrix allGlue}
\end{align}
which can be written as a polynomial in $\Nc$ using \eqref{eq:colour algebra}.

Similarly, the amplitude with a single quark line is given by\footnote{
We consider all particles as outgoing, 
so each quark line has a quark and an antiquark.}
\begin{equation}
\mathcal{M}(q\bar{q}+ng) = \sum_{P(1,\dots,n)}(t^1\dots t^n)_{q\bar{q}}M(1,\dots,n)~,
\end{equation}
with colour matrix $C^{q\bar{q}+ng}_{\si\si'}$
\begin{equation}
C^{q\bar{q}+ng}_{\si\si'} = (t^{\si_1}\dots t^{\si_n} )_{q\bar{q}}(t^{\si'_n}\dots t^{\si'_1})_{\bar{q}q}~,
\label{eq:col matrix 2q}
\end{equation}
% AL 220810: checked LC terms against MG implementation
while the amplitude with two distinct quark lines is given by

\begin{align}
\mathcal{\hat{M}}(q\bar{q}Q\bar{Q}+ng) &= 
\sum_{i=0,n}\sum_{P(1,\dots,i)}\sum_{P(i+1,\dots,n)} 
\left[\vphantom{\frac{1}{\Nc}}\right. \nonumber \\
&(t^1\dots t^i)_{q\bar{Q}}(t^{i+1}\dots t^n)_{Q\bar{q}}\nonumber \\
&\quad \times M(q,1,\dots,i,\bar{Q},Q,i+1,\dots,n,\bar{q}) \nonumber \\
&-\frac{1}{\Nc}(t^1\dots t^i)_{q\bar{q}}(t^{i+1}\dots t^n)_{Q\bar{Q}}\nonumber \\
&\left.\vphantom{\frac{1}{\Nc}}\quad\times M(q,1,\dots,i,\bar{q},Q,i+1,\dots,n,\bar{Q})\right]~.
\label{eq:4q amplitude fundamental}
\end{align}
% AL 220810: checked LC terms against MG implementation
Here, the first sum allows the gluons to be emitted by either fundamental colour line,
and the second and third sums permute the gluons on each fundamental colour line.
If there are no gluons in a string of $t$-matrices ($i = 0$ or $i=n$),
then that string should be replaced by a Kronecker delta with the relevant
(anti)fundamental indices.

The reason to have two strings of $t$-matrices,
is that we have used the Fierz identity (last equation of \eqref{eq:colour algebra})
to remove the repeating colour index of the internal gluon connecting the two quark lines.
This leaves us with two terms, the first (second line of \eqref{eq:4q amplitude fundamental})
is called the $\un$ term, while the second (fourth line of \eqref{eq:4q amplitude fundamental})
is called the $\uone$ term, and is $1/\Nc$ suppressed. 

% In \mgs{} the $\un$ and $\uone$ terms in \eqref{eq:4q amplitude fundamental}
% are treated as separate colour orderings, and the factor $-1/\Nc$
% is absorbed into the kinematic amplitude $M$.
% \commentAL{Put the words $\uone$ and $\un$ gluon here somewhere so later on it makes sense!}

If the two quark lines have the same flavour we use (see e.g.\ \cite{Weinzierl:2005dd,Reuschle:2013qna})
\begin{align}
\mathcal{M}(q_1\bar{q}_1q_2\bar{q}_2+ng) &=
\mathcal{\hat{M}}(q_{\si(1)}\bar{q}_1q_{\si(2)}\bar{q}_2 + ng) 
\nonumber \\
&- 
\mathcal{\hat{M}}(q_{\si(2)}\bar{q}_1q_{\si(1)}\bar{q}_2 + ng)~,
\label{eq:4q amplitude same flavour}
\end{align}
where $\si$ is a permutation of the quarks, and $\mathcal{\hat{M}}$ 
is the distinct-flavour amplitude from \eqref{eq:4q amplitude fundamental}.

\subsubsection{$1/\Nc$ Expansion}
\label{sec:1/nc expansion}
% \OMsout{Here we detail the $1/\Nc$ colour expansion used in this paper.}
In the fundamental basis, each term in the colour matrix $C_{\si\si'}$ 
is a polynomial in $\Nc$. 
One possible definition of the colour expansion in this basis, 
is to keep polynomials of the highest degree at leading colour (LC),
keep polynomials with at most two degrees smaller at next-to-leading colour (NLC),
and so on.
In this definition, each kept polynomial is retained in full, 
i.e.\ we do not truncate the individual polynomials in the colour matrix.
We now go through the expansion for different types of tree-level amplitudes.
% \AL{Rephrase this such that is theoretical definition choice, not an implementation choice }
% In theory could've truncated the polynomials in the expansion as well and been accurate to the given colour order}
% \AL{Move to near \eqref{eq:col matrix single row} to also explain that no jamps are repeated in subsequent colour orderings}

\paragraph*{Amplitudes with at most one quark line:}
~For these amplitudes, the polynomial has the form \cite{Mangano:1990by}
% \AL{Were we sure that the expansion here always decreases by two powers of $\Nc$?
% Mangano + Parke  appear to say so.
% }
\begin{align}
C_{\si\si'} &= a_n\Nc^{n} + a_{n-2}\Nc^{n-2} + \dots + a_m\Nc^m~,
\nonumber \\
\text{for }&\begin{cases}
n = n_g, & \text{all-gluon amplitudes }\\
n = n_g+1, & \text{single-quark amplitudes}
\end{cases}
\label{eq:col expansion < 1 q line}
\end{align}
where each term in the expansion is two powers of $\Nc$ smaller than the previous term,
each $a_{i}$ is some constant, $n_g$ is the number of gluons and $m$ is an integer with $m \le n-2$.
This motivates expanding the colour matrix in powers of $\Nc$,
such that the LC terms are those with $a_n \neq 0$,
the NLC terms are those with $a_n = 0, a_{n-2} \neq 0$, 
and so on. 

Looking at the colour matrices themselves (\eqsrefa{eq:col matrix allGlue}{eq:col matrix 2q}), 
and using the colour algebra relations
\eqref{eq:colour algebra},
it is easy to prove that $a_n = 0$ only if $\si \neq \si'$, $a_{n-2} = 0$ 
except on the diagonal and some off-diagonal terms, and so on.
% \OM{skip below?}
% This means that if we want a colour expansion accurate to some $\mathrm{N^kLC}$, 
% we require the following parts of the colour matrix:
% \begin{itemize}
% \item LC: Only the diagonal terms $C_{\si\si}$
% \item NLC: Only diagonal terms $C_{\si\si}$ and off-diagonal terms $C_{\si'\si}$
% whose leading power of $\Nc$ is $\Nc^{n-2}$
% \item NNLC: off-diagonal terms $C_{\si\si'}$ whose leading power of $\Nc$
% is $\Nc^{n-4}$, as well as NLC and LC terms
% \item etc.
% \end{itemize}

\paragraph*{Modified leading colour for all-gluon amplitudes:}
~The LC all-gluon amplitude can be modified and made more accurate by using \cite{Mangano:1987xk,Mangano:1990by}
\begin{align}
\sum_{\mathrm{colours}}|\mathcal{M}(ng)|^2 &= \TR^n\Nc^{n-2}(\Nc^2-1)\nonumber \\
&\sum_{P(2,\dots,n)}\Big[|M(1,\dots,n)|^2 + \mathcal{O}(\Nc^{-2})\Big]~,
\label{eq:allGlue mod LC}
\end{align}
as the LC definition. 
Note that in this definition we do not keep the full LC polynomial, 
but rather truncate it due to relations between colour-ordered amplitudes.

% \ALremove{As we will see in \secref{sec:colour accuracy},
% using this approximation is extremely beneficial}. 
Unfortunately, the authors only know the $\mathcal{O}(\Nc^{-2})$
terms in this version of the expansion for 6 or less gluons \cite{Mangano:1987xk},
but not in full generality. 
This leads to the strange effect that the default `leading colour' amplitude 
\eqref{eq:allGlue mod LC} is more accurate than the NLC amplitude which uses the standard $1/\Nc$
expansion in the fundamental basis (cf \secref{sec:colour accuracy}). 
For this reason, we label the default LC matrix element as modified LC, or `modLC'. 
We leave to future work a program which calculates the modified off-diagonal terms for an arbitrary number of gluons. \\

\paragraph*{Amplitudes with two quark lines:}
~The colour expansion for these amplitudes suffers from two problems,
one which occurs when the quarks have the same flavour,
and another when they have distinct flavours.
First, unlike \eqref{eq:col expansion < 1 q line},
the same-flavour colour matrix has the form
\begin{align}
    C_{\si\si'} &= a_n\Nc^{n} + a_{n-1}\Nc^{n-1} + a_{n-2}\Nc^{n-2} + \dots + a_m\Nc^m~,
    \nonumber \\
    n&=n_g+2~,
    \label{eq:col expansion 2q lines}
\end{align}
so that, at a given order of the expansion, 
we have corrections of $\mathcal{O}(1/\Nc) \sim 0.33$,
not of $\mathcal{O}(1/\Nc^2) \sim 0.11$ as before.
Due to this, we do not expect as precise an expansion as the previous cases.

% so that different colour factors can be separated by one power of $1/\Nc$, not two as before.
% This means that, at a given order of the expansion, 
% we have corrections of $\mathcal{O}(1/\Nc) \sim 0.33$,
% not of $\mathcal{O}(1/\Nc^2) \sim 0.11$ as before.
% Due to this, we do not expect as precise an expansion as the previous cases.

Despite this, we still define the expansion in powers of $1/\Nc^2$, and not as powers of $1/\Nc$. 
Therefore, the LC terms are those with $a_n \neq 0$,
the NLC terms are those with $a_{n-1} \neq 0$ and/or $a_{n-2} \neq 0$ but $a_n=0$,
and so on.

% That is, at any order in the expansion, we have corrections of $\mathcal{O}(1/\Nc) \sim 0.33$,
% not of $\mathcal{O}(1/\Nc^2) \sim 0.11$ as before.
% Due to this, we do not expect as precise an expansion as the previous cases.
% Note that despite this,
% we still define the expansion in terms of powers of $1/\Nc^2$, 
% \textit{not} powers of $1/\Nc$. \OM{be more clear what is LC and what is NLC here.}

If the quarks have distinct flavours, the colour matrix once again follows \eqref{eq:col expansion < 1 q line} with $n = n_g+2$,
but this time a different problem arises.
In this case, 
at LC we only include the first three lines of \eqref{eq:4q amplitude fundamental},
missing entirely all of the kinematic amplitudes in the last line of this equation.
Since these kinematic amplitudes could contain terms much larger than $1/\Nc^2$
we expect the expansion to be poor at LC.
In \appref{sec:modLC multiquark} we show an attempt to solve this second problem
by redefining the colour expansion.

\subsection{Berends-Giele Recursions}
\label{sec:BG recursion}
% \begin{itemize}
%     \item Probably good to mention that implementing BG recursion is not new and cite where people have already done this
%     \item Perhaps give formulae for all-gluon and for quark line
%     \item Perhaps give schematic example similar to MCnet talk (but correct!!) 
%     to demonstrate difference compared to old algorithm
% \end{itemize}

The basic idea of these recursions is to calculate an off-shell current
$J_n(1,\dots,n)$ with $n$ particles on shell and a single particle off shell.
The $(n+1)$-particle colour-ordered amplitude is given by $J_n$ 
with its off-shell propagator amputated,
and the result contracted with the wavefunction for particle $n+1$ \cite{Berends:1987me}.

\paragraph*{Gluon currents:}
~The base ingredients of the gluon off-shell currents are the one- and two-particle currents $J_1^\mu$ and $J_2^\mu$
\begin{align}
J_1^{\mu}(1) &= \epsilon^\mu(1)~,
\nonumber \\
J_2^\mu(1,2) &= \frac{-i}{(p_1+p_2)^2}
V_3^{\mu\mu_1\mu_2}(p_1,p_2)J_{1,\mu_1}(1)J_{1,\mu_2}(2) ~,
\label{eq:1 and 2 gluon currents}
\end{align}
% AL 20220831: propagator same as in chi-flow = Feynman gauge propagator, all conventions for vertex can be reasonably absorbed in V_3
where $\eps^\mu(1)$ is the gluon polarisation vector with momentum $p_1$, and
$V_3^{\mu\mu_1\mu_2}(p_1,p_2)$ the colour-ordered three-gluon vertex.

Using these ingredients as input, 
together with the colour-ordered four-point vertex $V_4^{\mu_1\mu_2\mu_3\mu_4}$,
a generic $n$-point current $J_n^\mu$ is 
\begin{align}
&J_n^\mu(1,\dots,n) = \frac{-i}{P_{1,n}^2}\times\nonumber \\
&\left\lbrace \sum_{i=1}^{n-1}V_3^{\mu\nu\rho}(P_{1,i},P_{i+1,n})
J_\nu(1,\dots,i)J_\rho(i+1,\dots,n) \  + 
\right.
\nonumber \\
 & \ \left.\sum_{i=1}^{n-2}\!\sum_{j=i+1}^{n-1}\!\!\! V_4^{\mu\nu\rho\sigma}\!
J_\nu(1,\dots,i)J_\rho(i+1,\dots,j) 
J_\sigma(j+1,\dots,n)\!\!\right\rbrace\!\!, 
\label{eq:n gluon current}
\end{align}
% AL 20220831: propagator checked and using Feynman gauge, rest checked for consistency, all factors of e.g. sqrt(2) absorbed into vague V_3,V_4 defn
where we use the shorthand $P_{1,n}^2 = (p_1 + \dots + p_n)^2$,
drop the number of particles $n$ in $J_n^\mu$ where convenient,
and use all outgoing momenta.

To obtain the $(n+1)$-point amplitude it remains to amputate the propagator,
and contract this current with an (on-shell) external gluon,
\begin{equation}
M(1,\dots,n+1) = iP_{1,n}^2\eps_{\mu}(n+1) 
J_n^{\mu}(1,\dots,n)|_{p_{1} + \dots + p_{n+1}=0}~.
\label{eq: n+1 gluon BG amplitude}
\end{equation}
% AL 20220831: checked equation for consistency by hand and by comparing to BG paper

\paragraph*{Quark currents:}
~The base ingredients for the quark current is a single on-shell quark,
and an on-shell quark which radiated a gluon i.e.\
\begin{align}
J_1(q) &= \ubar(q)~, \nonumber \\
J_2(q;1) &= \ubar(q)i\gamma^\mu\eps_\mu(1) i \frac{\slashed{q}+\slashed{p}_1+m}{(q+p_1)^2-m^2}
\nonumber \\
&\equiv -J_1(q)\slashed{J}_1(1) \frac{\slashed{q}+\slashed{p}_1+m}{(q+p_1)^2-m^2}~,
\end{align}
% AL 20220831: checked for consistency with Feynman rules and with BG paper
% no g_s because then we'd have to include it recursively and this doesn't work.
% Only required in the amplitude
where if the current $J$ has a $q$ in its arguments, then it is a quark current,
 otherwise it is a gluon current.

For an arbitrary number of gluons, the quark current is
\begin{align}
J(q;1,\dots,n) = -\sum_{i=0}^{n-1}&J(q;1,\dots,i)\slashed{J}(i+1,\dots,n)\nonumber \\
 &\times  \frac{\slashed{q}+\slashed{P}_{1,n}+m}{(q+P_{1,n})^2-m^2}~,
\end{align}
% AL 20220831: compared to eq. (3.9) BG paper
while the amplitude is found by contracting with the inverse propagator
and the antispinor, and putting the anti-spinor on shell 
\begin{align}
M(q;1,\dots,n;\qbar) = &J(q;1,\dots,n) (-i) (\slashed{q}+\slashed{P}_{1,n} -m)
\nonumber \\
&\times v(\qbar)|_{q+P_{1,n}+\qbar = 0}~,
\label{eq:quark amplitude from bg}
\end{align}
% AL 20220831: compared to previous equations and to eq. (4.10) BG paper
where again $P_{i,j} = p_i+\dots+p_j$ and all momenta are outgoing.

% We can similarly use an anti-quark to define the spinor current, 
% with base elements
% \begin{align}
% J_1(\qbar) &= v(\qbar)~, \nonumber \\
% J_2(1;\qbar) &= i \frac{-\slashed{\qbar}-\slashed{p}_1+m}{(\qbar+p_1)^2-m^2} i\gamma^\mu\eps_\mu(1) v(\qbar)\nonumber \\
% &\equiv  - \frac{-\slashed{\qbar}-\slashed{p}_1+m}{(\qbar+p_1)^2-m^2} \slashed{J}_1(1) J_1(\qbar)~.
% \end{align}
% \AL{Is this $+m$ or $-m$?? We don't change the propagator numerator because we started with a v spinor right??
% I think the BG paper seems to have $\slashed{p}-m$ but $p < 0$ so all positive and no minus sign in the front.
% Whichever it is, propagate correct solution to next two equations}
% A generic anti-quark current is therefore
% \begin{align}
% J(1,\dots,n;\qbar) &= - \frac{-\slashed{\qbar}-\slashed{p}_1+m}{(\qbar+p_1)^2-m^2}
% \\ \nonumber 
% &\times\sum_{i=0}^{n-1}\slashed{J}(1,\dots,i)J(i+1,\dots,n;\qbar)
% ~,
% \end{align}
% with the amplitude found similarly to as in \eqref{eq:quark amplitude from bg}
% \begin{align}
% M(q;1,\dots,n;\qbar) &= \bar{u}(q)(-i) (-\slashed{\qbar}-\slashed{P}_{1,n} -m) \\ \nonumber
% &\times J(1,\dots,n;\qbar)|_{q+P_{1,n}+\qbar = 0}~.
% \end{align}

\section{Technical Implementation}
\label{sec:implementation}
In this section, we will go through some of the details of our implementation of the colour matrix and its expansion,
as well as of the BG recursions.
First, in \secref{sec:mg},
we recall the main features of the event generator used throughout the paper, \mg{} (\mgs{}). 
Then, we will give some details of how we implemented the colour expansion 
(\secref{sec:col expansion implementation})
and the BG recursions (\secref{sec:BG recursion implementation}).
Finally, in \secref{sec:sources of speed differences},
we discuss in detail the sources of speed difference between the old and new codes using 
$g g \rightarrow 5g$ as a test case.

\subsection{The \mg{} Event Generator}
\label{sec:mg}
% \begin{itemize}
%     \item Mention the idea of \mg. I.e.\ from model to output to evaluate
%     \item Say that there are two main outputs of \mg, standalone and madevent. 
%     Explain what each does and is for. Confirm we will focus on standalone
%     \item Perhaps give a demonstration of the algorithm of recycling propagators to calculate diagrams
%     \commentAL{From Olivier: Not sure if really needed, but can possibly recycle from Kiran's paper (the pseudocode)
%     and/or figure 2}
% \end{itemize}

\mgs{} is a metacode which writes a program in the user's preferred language to calculate either the squared matrix element 
(standalone mode) or cross section/event generation (\me{} mode) of a chosen process within a chosen model at either leading order (LO) or next-to-leading order (NLO).
For example, choosing the default language of Fortran, the default model of the Standard Model (SM),
and $gg\rightarrow gg$ at LO as a process,
\mgs{} will first generate the four Feynman diagrams in this process,
then write a Fortran program which either calculates its squared matrix element or cross section.
The user then runs the program to get their result. 

The most common usage of \mgs{} (at LO) is the \me{} mode, 
which returns the cross section for a given process, 
including all cuts required to compare to experiment. 
This requires both calculating the hard matrix element, 
and sampling phase space efficiently to obtain an accurate cross section.

On the other hand, the standalone version of \mgs{}
calculates matrix elements at a specific, given, phase-space point.
It allows to isolate the speed of a matrix element computation,
since we do not have to worry about the convergence speed of the integral.
% \OMsout{It is also most useful for confirming the accuracy of a matrix-element calculation.} 
If many phase-space points are required,
it uses RAMBO \cite{Kleiss:1985gy} to do a flat scan of the phase space.

In this paper, we use the standalone version to better isolate the speed of the matrix element calculation
and to validate that the Berends-Giele recursions are correctly implemented.
% This means that we use a flat scan of phase space to obtain the average accuracy of the $1/\Nc$ expansion. \OM{would remove this last sentence}

\subsection{Implementation of Colour Computation}
\label{sec:col expansion implementation}

% \begin{itemize}
%     \item Say how colour sum done in std \mgs{}
%     \item Mention single row of colour sum for BG plus permute \jamp{} number when permuting rows
%     \item Introduce the idea of different flows?
% \end{itemize}

% \commentAL{Use pseudocode to get point across!}

In standard \mgs{} (also referred to below as the old code), the colour matrix is written explicitly as a square matrix of floats with size growing factorially with the particle multiplicity, 
and \eqref{eq:colour sum} is calculated by using two for loops to do the explicit matrix
multiplication.
All of the Feynman diagrams appear on equal footing, 
and are only calculated once each using the helicity amplitude formalism.
In pseudocode, it looks like this:

\begin{lstlisting}[language=Bash]
 # store colour matrix
 C_ij = ...
 
 # calculate all wavefunctions and propagators
 wf_i = ...
 
 # calculate all Feynman diagrams
 amp_i = ...
 
 # combine Feynman diagrams into colour-ordered amplitudes
 coamp_i = ...

 # sum over colours
 full_amp^2 = 0
 loop over rows
 |  tmp = SUM over columns (C_ij * coamp_j)
 |  full_amp^2 += conj(coamp_i) * tmp
\end{lstlisting}

In the new BG/colour ordered code (from now on referred to as the new code) each of these steps are done differently.
One big difference occurs for multiquark amplitudes.
For these, we do not simply calculate all kinematic (BG or Feynman) diagrams once.
Instead, we separate the kinematic amplitude into multiple calculations of different flows corresponding to:
(i) whether the colour ordering belongs to a \un\ or \uone\ gluon;
and (ii) how many gluons are on each colour line.
This makes it easy to combine partial graphs into BG currents, 
but has the disadvantage that the same kinematic diagrams are calculated multiple times.

Additionally, instead of writing the full colour matrix explicitly,
we take the first row (if multiquark the first row for each flow) of the colour sum and separate it into contributions at LC, NLC, N2LC, etc.
% \AL{do we need to define N2LC here or is it obvious after defining NLC etc.? }
For each colour order and flow, 
we write the kinematic amplitudes for that row times the relevant colour matrix entries times the conjugate amplitude.
That is, we have something of the form
% \begin{align}
% \tet{jamp[1]*conj(c[1,i]*jamp[i] + c[1,j]*jamp[j] + $\dots$)}
% \label{eq:col matrix single row_old}
% \end{align}
\begin{equation}
M^*_{\sigma_1}\left(\sum_{j\in \text{N$^k$LC}}C_{\sigma_1\sigma_{j}}M_{\sigma_j} \right)~.
\label{eq:col matrix single row}
\end{equation}

To loop over all rows, we keep the values of colour factors in 
\eqref{eq:col matrix single row} the same,
and permute the colour-ordered amplitude indices $\si_j$. 
This requires a permutation matrix of the same size as the original colour matrix,
but which, unlike the colour matrix,  has integer components,
so uses only half the size in memory (and it can technically be reduced even further).
This is a feasible solution for the multiplicities we wish to probe,
but for higher multiplicities the factorial-squared growth will quickly become a problem
(expanding in powers of $1/\Nc$ offers one possible solution depending on the accuracy desired).
% \commentAL{Somehow and somewhere comment that the $n!
% \times n!$ matrix is still ok at our multiplicities but not for higher ones}
%\OM{comment that the expansion in term of LC/... offers a solution}

A pseudocode of the new program (for a given flow) is:
% \begin{lstlisting}[language=Bash]
%  # store permutation matrix
%  perm_matrix[i,j] = k
 
%  # calculate all colour-ordered amplitudes, possibly in different flows
 
%   # calculate all wavefunctions and propagators
%   wf_i = ...
 
%   # Calculate all Feynman diagrams
%   amp_i = ...
 
%   # combine Feynman diagrams into colour-ordered amplitudes
%   coamp_i = ... # linear combination of the amplitude
    
%  result=0
%  loop over colour orders required
%   ztemp = 0
%   loop over rows of colour permutation matrix
%      loop over jamps from different flows required
%       ztemp += coamp[perm_matrix[nrow,i]] * c_ij * conj(coamp[perm_matrix[nrow,j]])
%   result += ztemp
% \end{lstlisting}

% \OM{what about this?}
\begin{lstlisting}[language=Bash]
# calculate kinematics
...

# store permutation matrix
perm[i,j] = k

# sum over colours
full_amp^2 = 0
loop over rows of permutation matrix (nrow)
| onerow = 0
| loop over column j, filtered for 1/Nc
| |  onerow += C_1j * coamp[perm[nrow,j]]
| onerow *= conj(coamp[perm[nrow,1])
| full_amp^2 += onerow
\end{lstlisting}

Note that in both methods of computation, one can use the fact that the colour-matrix is symmetric to further optimise the computation.\footnote{
This optimisation will be added to \mgs{} version 3.5.0 within the old code and within 3.5.1 for the new code.}

\subsection{Implementation of Berends-Giele Recursion}
\label{sec:BG recursion implementation}

\begin{figure}
\resizebox{0.5\textwidth}{!}{%
  \includegraphics{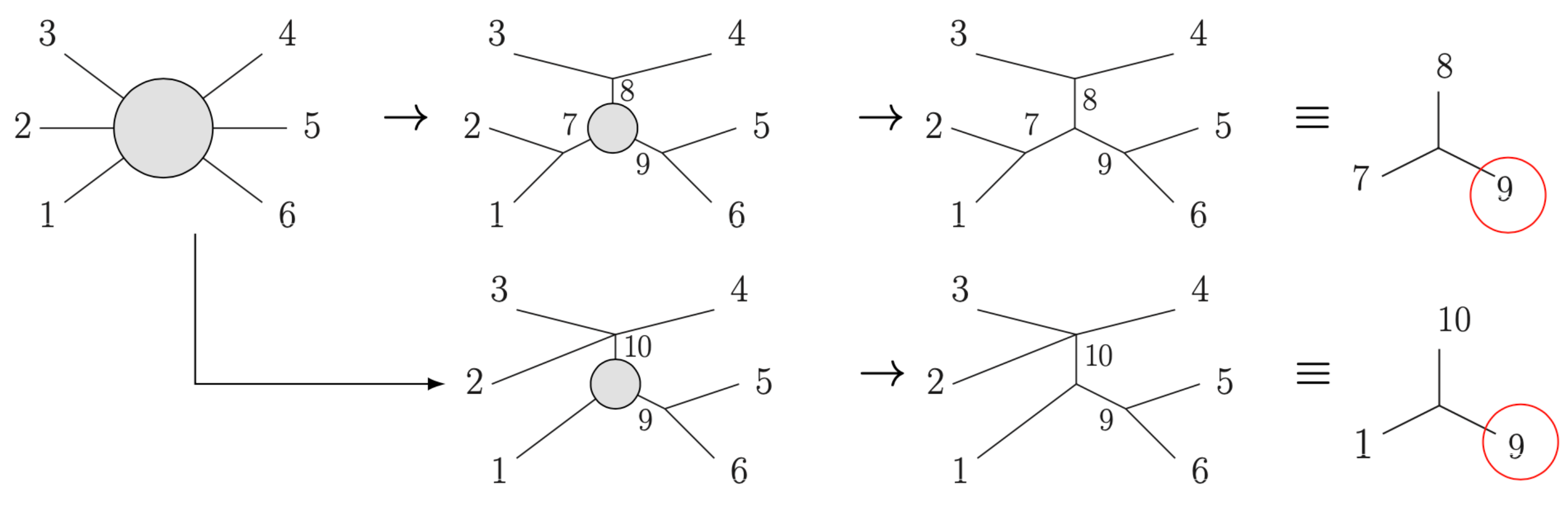}
}
\caption{Recycling in \mgs{}.
Starting from out to in, \mgs{} calculates propagators as off-shell currents which it then caches.
When two diagrams share an off-shell current such as current 9, 
there is no need to recalculate this current, and the cached version is instead used.
The value of the Feynman diagram is then calculated by contracting the remaining, often internal, currents.
}
\label{fig:mg recycling}       % Give a unique label
\end{figure}

In \mgs{}, multiple Feynman diagrams are calculated efficiently by recycling three- and four-point off-shell currents when they belong to multiple diagrams
(see \figref{fig:mg recycling}).
This allows to reduce the total number of calculations required,
making a simpler and faster program.

\begin{figure}
\resizebox{0.5\textwidth}{!}{%
  \includegraphics{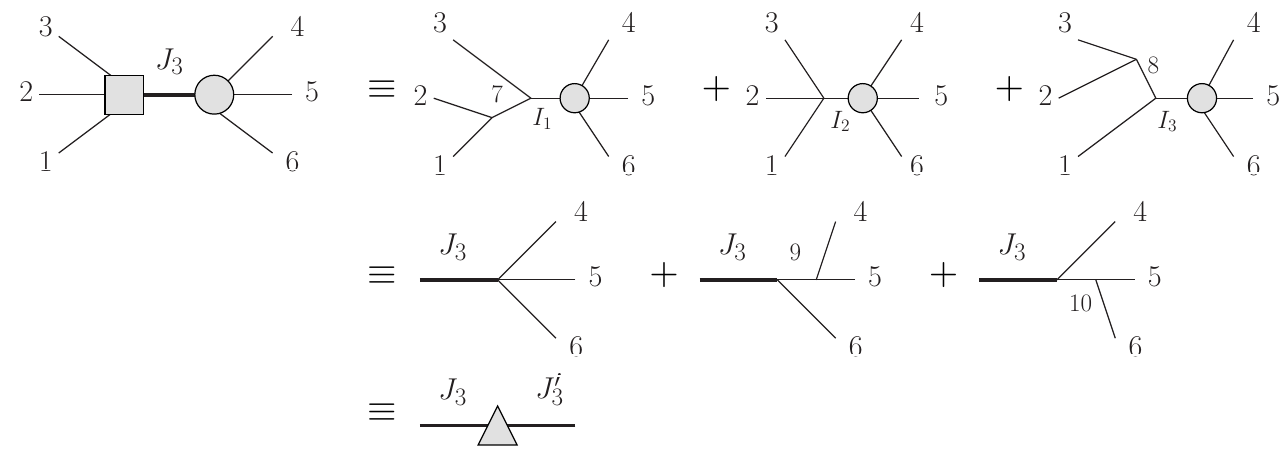}
}
\caption{An example of BG recursion in \mgs{}.
Each of the square and circular blobs represent three possible diagrams.
The current $J_3$ is created by combining the three off-shell currents 
$I_1,I_2$, and $I_3$ into a single off-shell current (first line),
which is then used in three graphs (second line).
In this way, 9 Feynman diagrams becomes 3 graphs.
A future optimisation would be to also combine particles 4, 5, and 6 into another three-particle current
$J_3'$, which would then have its propagator amputated and be contracted with $J_3$ (third line). 
}
\label{fig:BG recursion}       % Give a unique label
\end{figure}

% \AL{Rikkert confused by this explanation, rephrase?
% Also Malin wasn't sure why we needed 6 or more particles, should probably rephrase}
While the version of BG recursion given in \secref{sec:BG recursion} 
builds currents by always adding a single extra particle until all particles have been used,
\mgs{} does not do this.
This is because \mgs{} uses multiple small BG currents in parallel (for different external particles),
before eventually contracting these currents together in a trivalent or four-valent vertex
(see first two lines of \figref{fig:BG recursion}).
One consequence of this choice is that BG recursions lead to a speed gain only for multiplicity greater than or equal to six, 
since below that the recycling algorithm reaches the same efficiency.
 
% we can only use BG recursion when we have six or more particles in the process
% \AL{(for 5 or less particles the Feynman diagram is completed before a BG current can be formed)}.
% % AL 220811: if 5 particles and trivalent vertices will always create two sets of 2->1 and then with remaining particles create 3->0 to get amp
% % if 6 particles and trivalent (and 4-valent) vertices can combine as in \figref{fig:BG recursion}
% Therefore, this algorithm can only be faster than standard \mgs{}
% for multiplicity greater than or equal to six.

We stress that our new code is less optimal to compute the kinematics part than standard \mgs{}, both with and without using BG recursion.
The reason for this is that we have not implemented all possible optimisations 
(many such optimisations are well known, and are left to future work).
Nevertheless, the BG recursions compile far quicker than the old code at high multiplicity, allowing  to generate and study processes with higher multiplicity than before.
Also,
the new code is faster to run than the old code at high multiplicities, even with the slower kinematics (see
\secref{sec:speed gain}).

% Therefore, if we keep the rest of the algorithm the same,
% we expect BG recursion to be faster than standard \mgs{} for multiplicity greater than or equal to six,
% but roughly the same speed at lower multiplicity.

% \OM{Should we mention here that we have a new way to generate the code with/without BG.
% That both of them actually produce less optimised code compare to standard madgraph.
% But that they allow to go higher in multiplicity?}

% \AL{we should stress that we are less optimal than MG with Feynman (but many optimisations known).
% We also have a huge advantage that we can do $2\rightarrow 6$ with BG but not without (stress that we suck, not BG sucks).
% See Olivier's comment at end of subsec}

\subsection{Sources of Speed Differences}
\label{sec:sources of speed differences}

As seen in the pseudocode in \secref{sec:col expansion implementation},
we can, loosely speaking, divide a \mgs{}
calculation into four parts:
% Loosely speaking, we can divide a \mgs{}
% calculation into four parts:
% \OM{Make reference to psuedo code (as shown in pseudo code etc.)}
\begin{enumerate}[label=(\roman*)]
    \item Calculate wavefunctions (\wfs), both external and internal (i.e.\ propagators or off-shell BG currents)
    \label{enum:speed wfs}
    \item Calculate the amplitudes (\amps), i.e.\ completed Feynman or BG graphs
    \label{enum:speed amps}
    \item Sum up the \amps{} into the colour ordered amplitudes ($M_\sigma$)
    \label{enum:speed jamps}
    % \jamps{} (i.e.\ $M_\sigma$)
    \item Loop over the colour matrix, calculating \eqref{eq:colour sum}.
    \label{enum:speed colsum}
\end{enumerate}
% In the new colour expansion with BG-like kinematics,
% all four of these steps are changed, 
% as described in 
% \secref{sec:col expansion implementation}.
% As an example, the effects of these changes for the process $g g \rightarrow 5g$
% are summarised in \tabref{tab:profile 7g}.
% \commentAL{Try to be more sure of col sum and $M_\si$ in \tabref{tab:profile 7g} before publishing!!
% also, put in other processes??}

% As described in the previous two subsections, 
% all four of these steps are changed in the new colour expansion with BG-like kinematics.
In the new code,
all four of these steps are changed.
To understand the effect of each change, we profiled the process $g g \rightarrow 5g$
for both standard \mgs{} and for the new code, with results summarised in \tabref{tab:profile 7g}.

\begin{table}
\caption{The number of instructions to calculate $g g \rightarrow 5g$ 
for 10 phase-space points at full colour in standard \mgs{}
standalone and in our new code (at N6LC, i.e.\ full colour and using BG recursions).
In addition to the total number of instructions required to do the calculation
(Full ME), 
we have broken down the calculation into four steps:
calculating internal and external wavefunctions (\wfs{}),
calculating completed graphs (\amps{}),
putting these graphs into colour ordered amplitudes ($M_\sigma$),
and summing over colours (col sum).
The number in brackets is the percentage of the total number of instructions required to calculate the Full ME.
In the right-hand column we compare the old code and the new one, 
and use red when the new code is worse than the old one.
% \commentAL{We have a factor 1.32 for number of instructions, and a factor of 1.24 in time spent on my laptop. Seems just about consistent.}\\
% \AL{call BG colour ordered or new code plus be consistent through paper}.
}
\label{tab:profile 7g}       % Give a unique label
\begin{tabular}{c|ccc}
\hline\noalign{\smallskip}
 & \mgs{} & new code & $\frac{\text{\mgs{}}}{\text{new code}}$ \\
\noalign{\smallskip}\hline\noalign{\smallskip}
Full ME & 41G  & 31G  & 1.3     \\
\ref{enum:speed wfs} \wfs &  0.99G $(2.4\%)$ & 0.59G $(1.9\%)$ & 1.7 \\
\ref{enum:speed amps} \amps & 6.6G $(16\%)$ & 16G $(51\%)$ & \textcolor{red}{0.42} \\
\ref{enum:speed jamps} $M_\sigma$ & 3.2G $(9.3\%)$ & 1.1G $(3.6\%)$ & 2.9 \\
\ref{enum:speed colsum} col sum & 30G $(72\%)$ & 14G $(44\%)$ & 2.2 \\
% \AL{Used 2 sig figs everywhere}\\
\noalign{\smallskip}\hline
\end{tabular}
\end{table}

For steps \ref{enum:speed wfs} and \ref{enum:speed amps},
our BG recursion misses many optimisations included in standard \mgs{},
so even though we use BG recursions,
we actually have more \wfs{} at low multiplicity but less at high multiplicity,
and have many more \amps{} in the new code than the old code.
Improving this is left for future work,
but for now we are mostly interested in high multiplicity processes where the colour sum dominates.
As seen in \secref{sec:speed gain} below,
at low multiplicity the missed optimisations cause the new program to be slower than the standard \mgs{} one,
but at high multiplicity the new program is significantly faster.
% At low multiplicity the missed optimisations causes the calculation to be slower than the standard \mgs{} one.

An effect of the BG recursions is to reduce how many \amps{} 
go into the individual colour-ordered amplitudes $M_\sigma$.
Though this part of the code was not a bottleneck, 
using BG recursions can improve this part of the calculation significantly at high multiplicity,
e.g.\ by about a factor three for $gg\rightarrow 5g$.

The biggest improvement of the new code is in the colour sum.
In standard \mgs{},
the colour matrix is stored as a 
matrix of real numbers of double precision. 
The colour sum is then just the matrix multiplication of \eqref{eq:colour sum}.

In contrast to this, 
the new code only explicitly stores the first row of the colour sum (for each flow).
We then have a single loop over all rows using a permutation matrix of integer numbers
(see \secref{sec:col expansion implementation} for more details).
This simple change appears to more than halve the work of the colour sum,
which is vital because as seen in \tabref{tab:profile 7g} and ref \cite{Mattelaer:2021xdr},
the colour sum in \mgs{} is one of the main bottlenecks for going to higher multiplicities.
While this change definitely helps, 
we remind that this optimisation doesn't change the factorial-squared growth of the colour sum. 
On the other hand, truncating the expansion in powers of $1/\Nc$
helps this issue.
% \AL{, especially if combined with phase-space symmetry \cite{Frederix:2021wdv}}.
%especially if one knows the relevant terms of the colour matrix in advance (as done in \cite{Frederix:2021wdv}).}
% \OM{mention here that truncating the expansion over 1/NC is solving such issue}

\section{Validation and Results}
\label{sec:validation and results}

Now we turn our attention to the results of this paper.
We will first look at the accuracy of the $1/\Nc$ 
colour expansion for various processes,
and validate this expansion by showing that it converges to the full colour result.
Next, we will consider the speed of the program and compare it with the standard version of \mgs{}.

We checked the accuracy and speed process by process in both pure QCD and mixed QCD/EW theories\footnote{
Note that the decay-chain syntax is not supported.},
with a representative subset of QCD processes shown below
(the mixed QCD/EW results are given in \appref{sec:EW results}).

As will be seen below,
the LC amplitudes are in general not good enough to be used in practical purposes,
the NLC amplitudes can be used to speed up phase-space integration but require special tricks/correction factors \cite{Alwall:2014hca,Danziger:2021eeg,Weinzierl:2000wd},
while all processes studied have good accuracy already at NNLC. 
For the speed, we will find that the new code is faster than the old one at high enough multiplicity, 
but slower for low multiplicities (where the computation is not dominated by the colour-matrix).

\subsection{Accuracy and Precision of Colour Approximation}
\label{sec:colour accuracy}
% \begin{itemize}
%         \item Show that the $1/\Nc$
%         expansion converges to the true result
%         \item Do a deep dive into the all-gluon results and comment on accuracy/precision/modLC
%         \item Show other 3 massless QCD procs in single figure and compare results to all gluon
%         \item Show other results here or in the appendix (e.g.\ $t\bar{t},w,z$ + QCD
%         \item Still comment on modLC for multiquark??
%         \item Conclude that NLC fairly precise but not necessarily accurate, N2LC is very precise and accurate for all studied processes
% \end{itemize}

\begin{figure}
\resizebox{0.5\textwidth}{!}{%
  \includegraphics{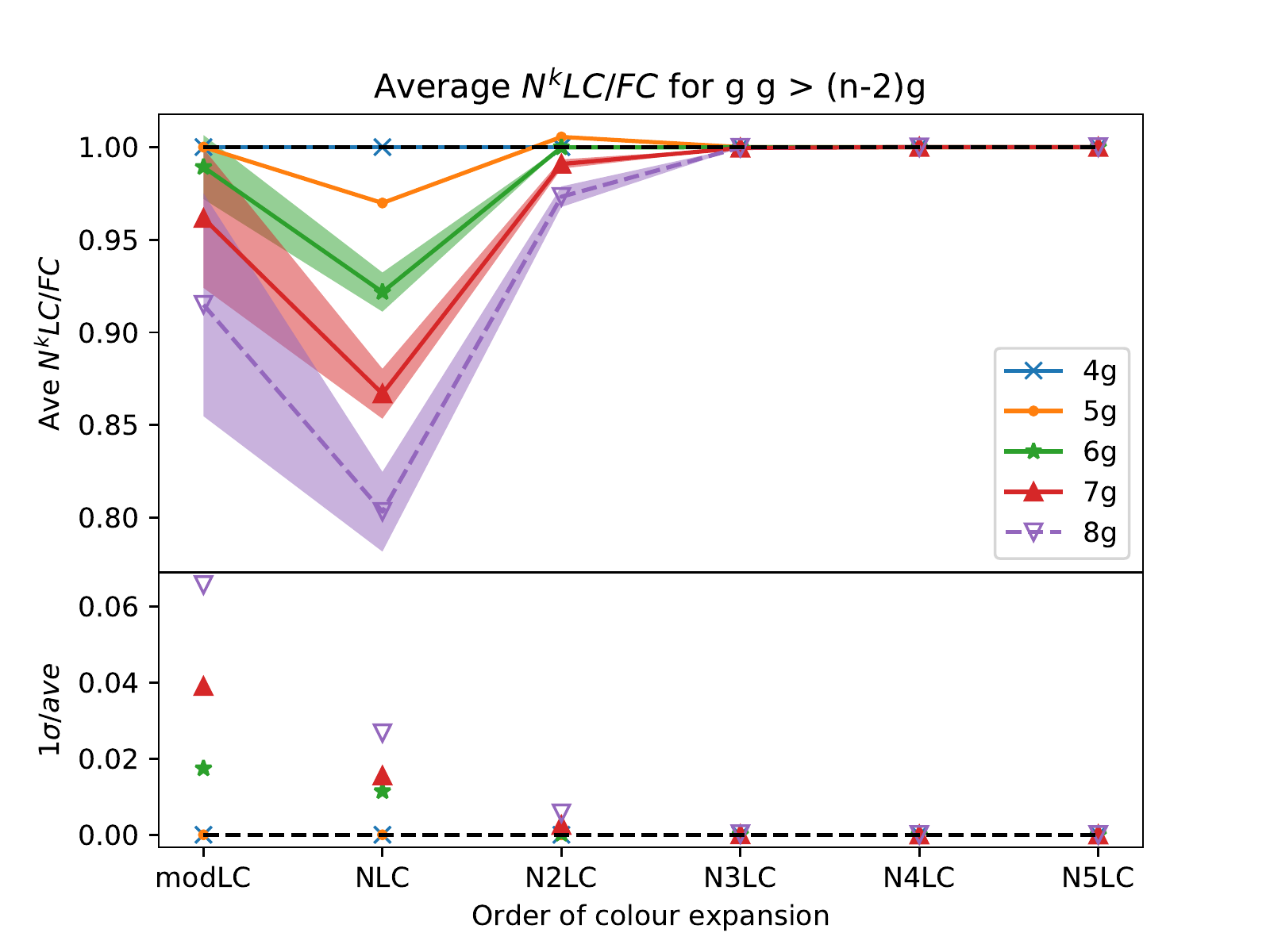}
}
\caption{Accuracy of expansion in colours $1/\Nc$ in the process
$gg \rightarrow (n-2)g$ compared to the value calculated in standard \mgs{} (FC).
modLC is described in \eqref{eq:allGlue mod LC} and is a modified LC value, 
hence it is more accurate than NLC which is unmodified.
The 8g line is dotted since the accuracy was compared to N5LC rather than FC due to \mgs{} not being able to calculate this process.
The top panel shows, for a given colour order N$^k$LC, the average N$^k$LC/FC value, with the standard deviation in the shaded region.
The bottom panel shows the relative error.
}
\label{fig:all-gluon accuracy}       % Give a unique label
\end{figure}

\paragraph*{All-gluon amplitudes:}
% \AL{Define $k$, or perhaps put it not as a power but as a normal letter so it's obvious that NkLC -> N2LC, N5LC etc.?}
~We begin by considering the accuracy of the $gg\rightarrow (n-2)g$
all-gluon amplitudes, as shown in \figref{fig:all-gluon accuracy}.
In the top panel we see the average value of N$^k$LC/FC
over a flat scan of phase space (using RAMBO \cite{Kleiss:1985gy}),
i.e.\ for each phase-space point we divide the colour-truncated squared matrix element by the full squared matrix element calculated by \mgs{}, 
and average this over phase space. 
% i.e.\ we average over phase space the squared matrix element with the truncated colour expansion divided by the full squared matrix element as calculated by \mgs{}. 
For up to 6 gluons, the average is taken using 100,000 phase-space points,
for 7 gluons using 10,000 points, and for 8 gluons using 1,000 points.
All processes were calculated at $\sqrt{s} = 1$TeV.
The 8g version is dotted since we could not compile the FC process in standard \mgs{},
therefore we took the N5LC value to approximate FC.
% in the denominator due to not having access to the full value from \mgs{}
% \OM{that the full colour is not possible to compile within MG5aMC standard code}.
Since the 8g N4LC and N5LC results already agree for the first four significant figures,
% Since the expansion converges to within a very good accuracy well before N5LC, 
this should not affect any conclusions. 
Such convergence is also a good validation of our colour expansion.

The shaded regions correspond to the standard deviation of the N$^k$LC/FC ratios,
while the bottom panel is the percentage uncertainty, 
i.e.\ the standard deviation divided by the average.
We assume a roughly Gaussian distribution\footnote{
We have made some basic checks that a Gaussian assumption is reasonable.
},
and study the phase-space dependence of the accuracy and precision later in this section.

% In calling these standard deviation and percentage uncertainty,
% we implicitly assume the error to be Gaussian and phase-space independent.
% We check the phase-space dependence for 
% \OM{comment on the gaussion hypothesis? and that we will study the phase-space dependence for other processes later?}

From \figref{fig:all-gluon accuracy},
we conclude that modified LC, \eqref{eq:allGlue mod LC},
is more accurate but less precise than NLC. Additionally,   
for 8 gluons the colour expansion converges by N3LC, at per-mil-level accuracy.
%AL 220811: 8 gluons at N3LC has about 0.1% accuracy. Smaller multiplicity is better
Also, by NLC the relative precision of the expansion is much smaller than the average offset from the true value,
allowing to systematically correct results if desired.
We stress that when computing cross-sections and/or generating events, 
precise but inaccurate results can help speed up the code.
This can be achieved by avoiding to compute the full matrix-element for all phase-space points, 
but still guaranteeing no bias after phase-space integration \cite{Krauss:2001iv,Alwall:2014hca}.

% \OM{This can also be done for NLC and following number presented in Table1}

To quantify the effect of modified LC, \eqref{eq:allGlue mod LC},
we show in \tabref{tab:all-gluon LC} 
the average values of both the standard LC/FC and the modified LC/FC. 
The NLC/FC value is also used for comparison,
confirming that it is far more accurate than the true LC amplitudes,
even if it is less accurate than the modLC results. 
Since the only difference between modLC and LC is changing the colour factor in \eqref{eq:allGlue mod LC},
the relative (but not absolute) precision of LC and modLC are the same.

\Tabref{tab:all-gluon LC} shows that a true LC all-gluon amplitude in the fundamental basis is a very poor description of the full amplitude,
being about 60\% too small for the 8 gluon amplitude.
The reason is likely that we are using fundamental matrices 
(i.e.\ the colour matrices of quarks)
to describe the colour of gluons. 
Therefore, we expect e.g.\ the colour flow expansion to be more suited to the all gluon amplitude, 
since a pure gluon amplitude can be fully described with $U(3)$
gluons.
Alternatively, the modLC description works very well since it uses more than just a strict expansion in colour to calculate the colour factor.

%The fact that,starting from NLC, the multi-gluon amplitudes are precise (even if not accurate) do offer opportunity when combine with a phase-space integration. Obviously one could monte-carlo over the inclusion of $N^kLC$ term for each given event. A better solution is use the equivalent of the virt-trick \cite{Alwall:2014hca}. In that method, we would first fit the ratio between the full color and the $N^kLC$ term (say $\alpha$) and then rewrite the integral as 
%\begin{equation}
%\int |M_{FC}|^2 = \int |M_{FC}|^2 -\alpha |M_{N^kLC}|^2 + \int \alpha |M_{N^kLC}|^2.   
%\end{equation}
%Then one can use importance sampling to limit the amount of time the full colour is needed while still avoiding any bias\footnote{The downside of this method is that it can leads to negative event. That effect can be kept under-control by under-estimating $\alpha$.}

\begin{table}
\caption{The average modLC/FC, LC/FC, and NLC/FC in the all-gluon colour expansion in the fundamental basis.}
\label{tab:all-gluon LC}       % Give a unique label
\begin{tabular}{c|ccc}
\hline\noalign{\smallskip}
 & modLC/FC & LC/FC & NLC/FC \\
\noalign{\smallskip}\hline\noalign{\smallskip}
4g & 1     & 0.704 & 1     \\
5g & 1     & 0.624 & 0.970 \\
6g & 0.989 & 0.549 & 0.922 \\
7g & 0.962 & 0.474 & 0.867 \\
8g & 0.915 & 0.401 & 0.803 \\
\noalign{\smallskip}\hline
\end{tabular}
\end{table}

\begin{figure}
\resizebox{0.5\textwidth}{!}{%
  \includegraphics{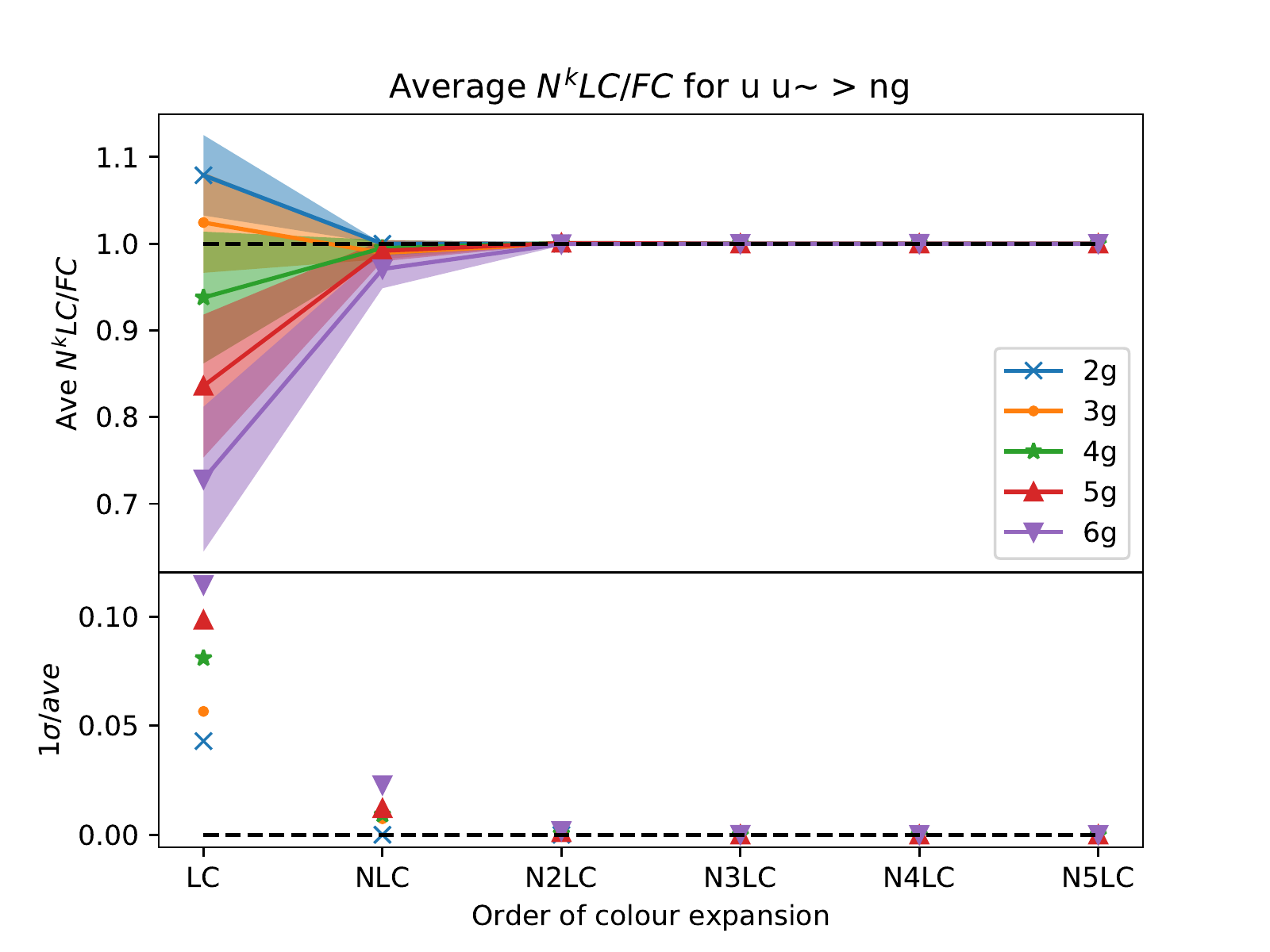}
}\caption{Same as \figref{fig:all-gluon accuracy} but for the process
$u\ubar \rightarrow ng$.}
\label{fig:quark accuracy}
\end{figure}

\paragraph*{Amplitudes with a single quark pair:}
~Next we consider QCD processes with a single quark pair using $u\ubar \rightarrow ng$ 
as a test process (see \figref{fig:quark accuracy}).
We used 100,000 phase-space points for up to 5 gluons and 10,000 points for 6 gluons.
In this case, the LC approximation is neither particularly accurate nor precise. 
At low multiplicity, it over-estimates the amplitude, 
while it increasingly under-estimates it starting from four gluon multiplicity.
Similar to the all-gluon amplitudes in \figref{fig:all-gluon accuracy},
the NLC relative precision is around a few percent.
However, unlike the all-gluon case,
the NLC amplitude is already quite accurate,
being on average percent-level accurate or better for 5 or less gluons,
and about 3\% accurate for 6 gluons. 
Both the accuracy and relative precision of N2LC is at or better than about 0.1\% for all studied processes.

\paragraph*{Amplitudes with two quark pairs:}
~To complete the pure massless QCD analysis,
we study processes with two quark pairs.
We take two test cases, $u\ubar \rightarrow d\dbar + ng$
and $u\ubar \rightarrow u\ubar + ng$, 
again using 100,000 phase-space points for all multiplicities except for the largest one, 
which was calculated using 10,000 points. 
We use two test cases here in order to study the effects of quark interference on accuracy and precision.

\begin{figure*}
\resizebox{1\textwidth}{!}{%
  \includegraphics{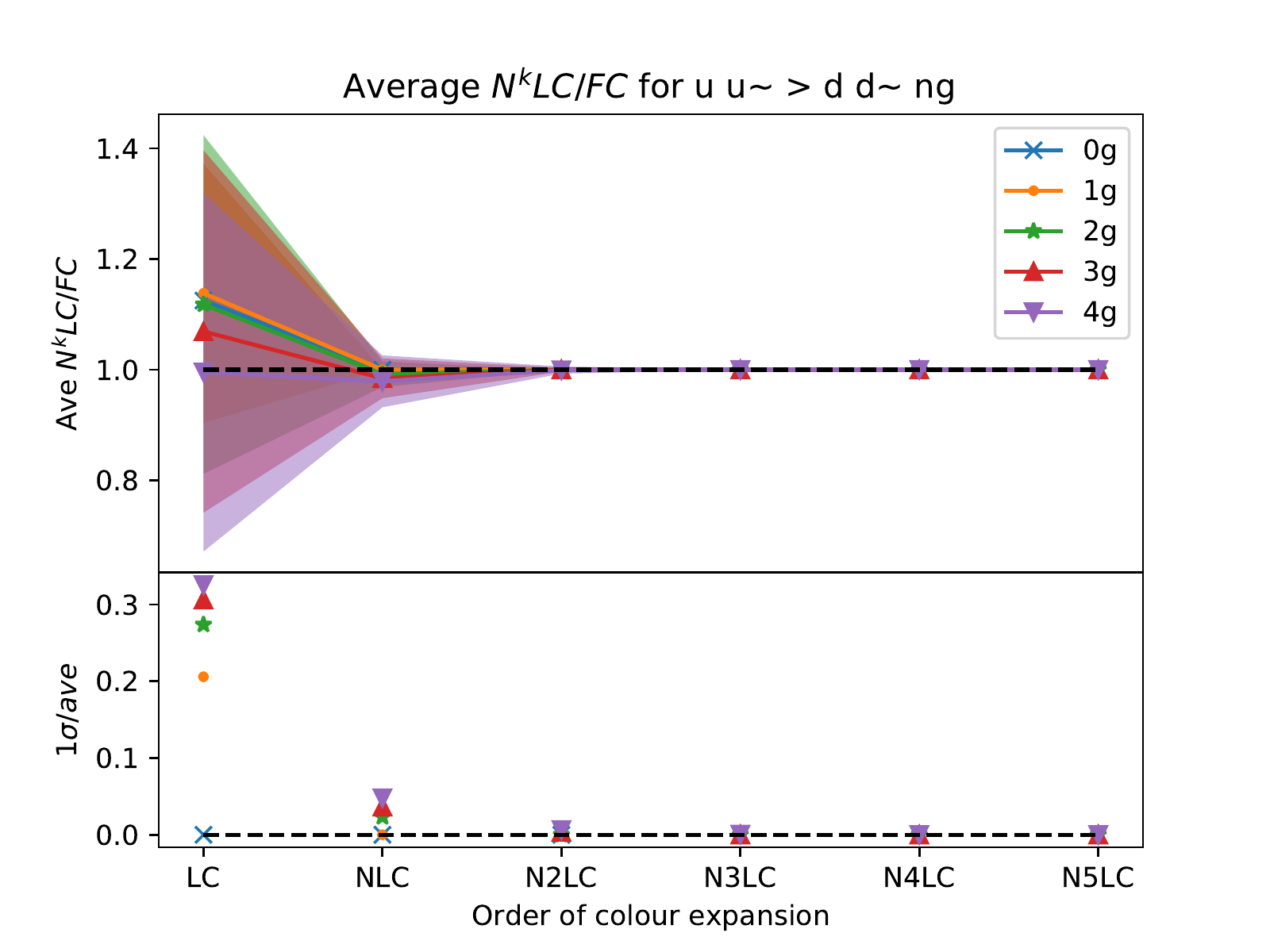}\hfill
  \includegraphics{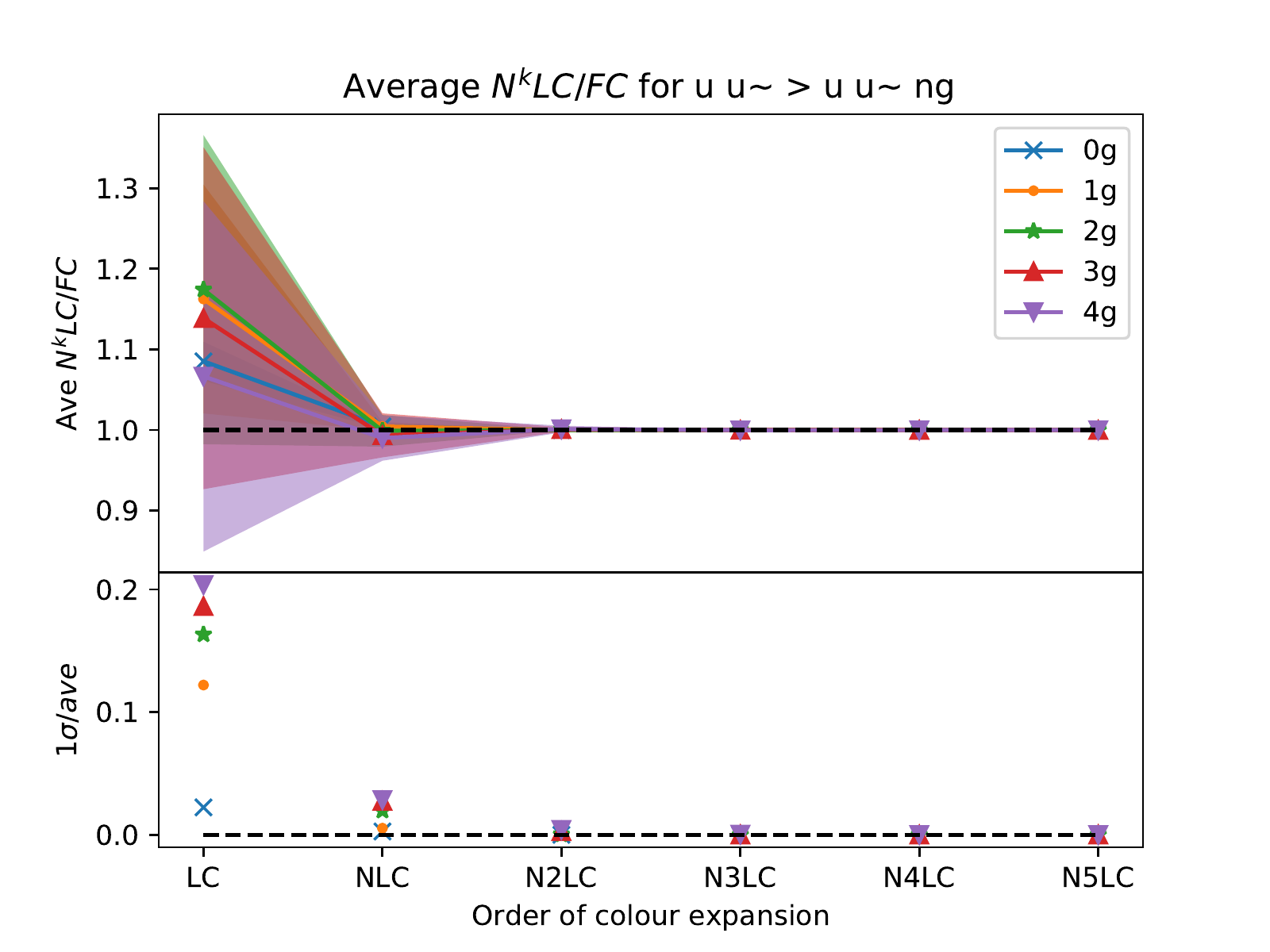}
}\caption{Same as \figref{fig:all-gluon accuracy} but for the processes 
$u\ubar \rightarrow d\dbar + ng$
and $u\ubar \rightarrow u\ubar + ng$.}
\label{fig:multiquark accuracy}
\end{figure*}

As we see in \figref{fig:multiquark accuracy},
LC has only about 20-30\% relative precision,
and that for distinct quark flavours the LC value again decreases with increasing gluons.
The same-flavour LC amplitudes are more precise than the distinct-flavour ones,
possibly due to all kinematic amplitudes being included already at LC for the same flavour case
(cf \eqsrefa{eq:4q amplitude fundamental}{eq:4q amplitude same flavour} 
and the discussion around \eqref{eq:col expansion 2q lines}). 
By NLC, the accuracy is already very good, 
around the percent level, 
with precision about 5\% or better.
Once again, by NNLC, the accuracy is around $0.1\%$ or better,
with precision around $0.5\%$ or better.

\begin{figure*}
\resizebox{1\textwidth}{!}{%
  \includegraphics{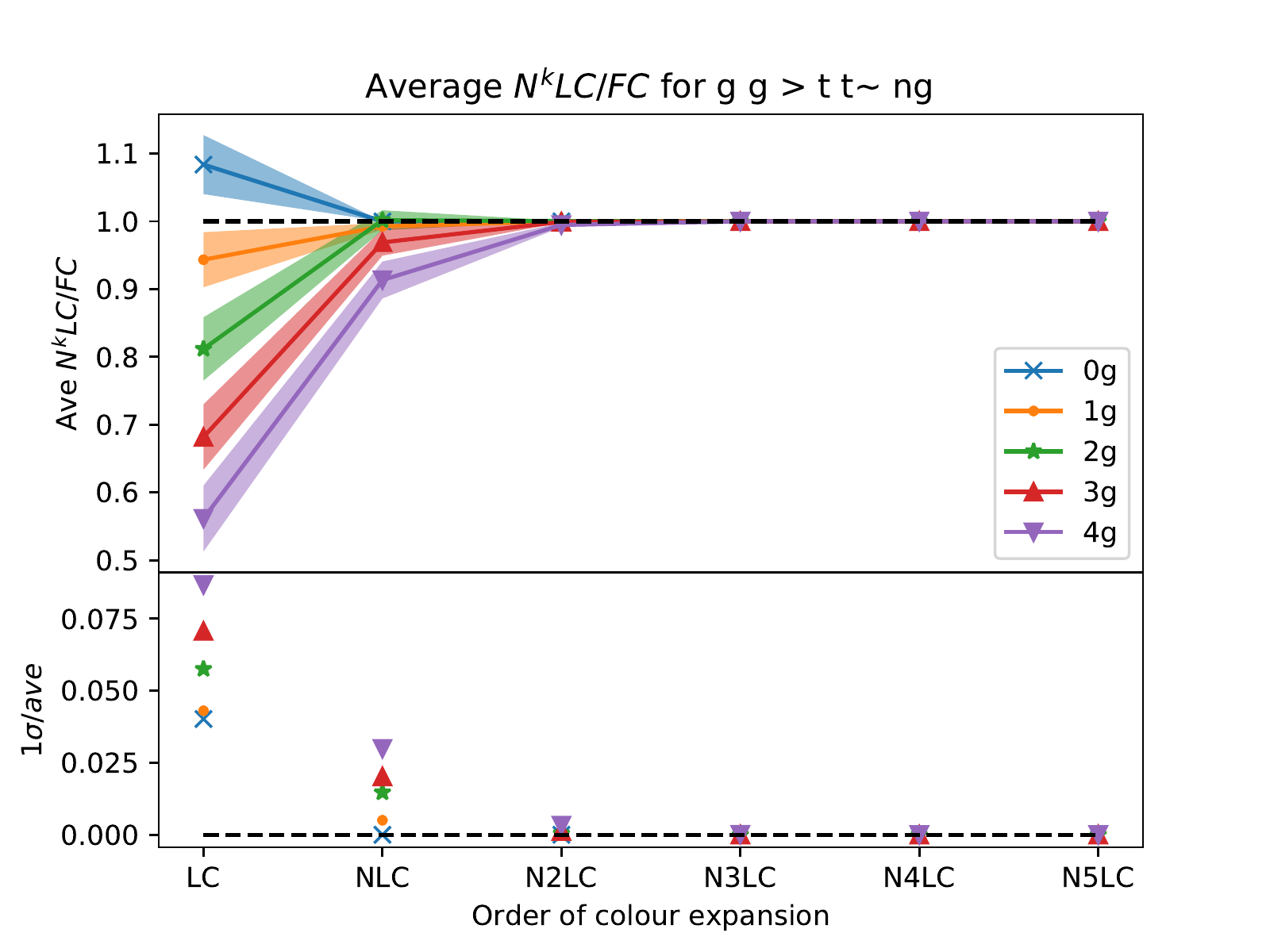}\hfill
  \includegraphics{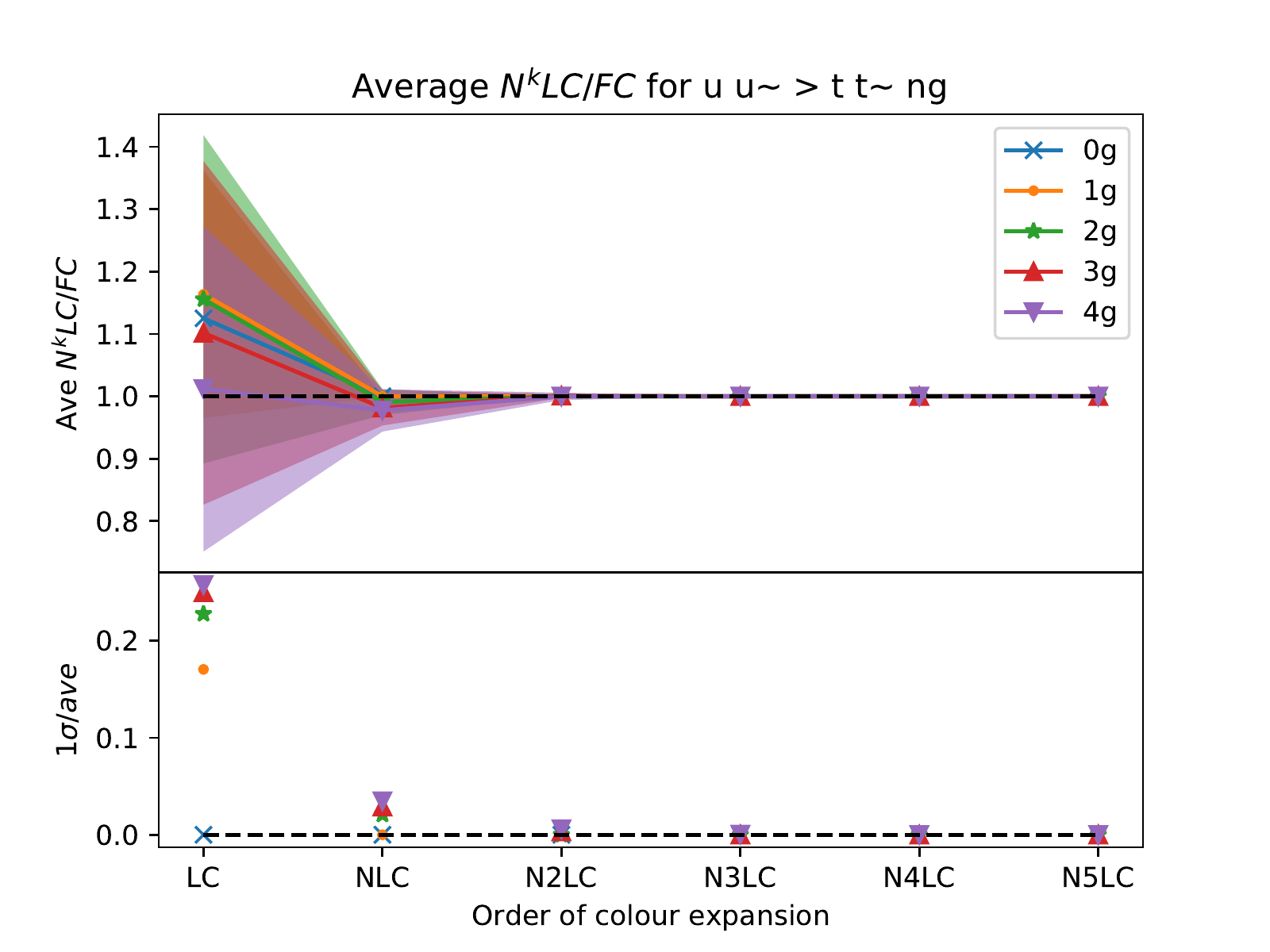}
}\caption{Same as \figref{fig:all-gluon accuracy} but for the processes 
$gg \rightarrow t\tbar + ng$
and $u\ubar \rightarrow t\tbar + ng$.}
\label{fig:ttbar accuracy}
\end{figure*}

\paragraph*{Amplitudes with a top quark pair:}
~An important process in QCD is the production of a top pair.
\mgs{} can now calculate this production using the new code.
As we see in \figref{fig:ttbar accuracy},
the LC values for $t\tbar$ production become very inaccurate at high multiplicity,
with the $gg\rightarrow t\tbar 4g$ matrix element being just $56\%$ of its required value on average.
However, the relative precision of about $8.7\%$ allows this value to be systematically corrected.
Indeed, such a correction for gluon-induced top production appears well motivated already for two or more final-state gluons.
If the process is quark-induced, 
the LC relative precision is around or above $20\%$ depending on the multiplicity.

At NLC, the results are also quite different depending on the subprocess.
For the gluon-induced process, 
the NLC result is only 9\% accurate for 4 gluons with a relative precision of about $2.9\%$,
while the second process is accurate to within a few percent for all processes studied but has a slightly worse relative precision of up to $3.5\%$. 

Like the previous processes, NNLC describes the results to a high accuracy and precision.
All processes are described to an accuracy and precision of about a half of a percent or better for all multiplicities.

\paragraph*{Accuracy in different parts of phase space:}
% \OM{NEED here to refer to section 6 of \cite{Mangano:1987xk}. They stated that the soft/collinear limit is encoded within all the colour-ordered amplitude which implies that we should not expect a particular pattern close to singularity. But it also state: "The 1/N term could destroy the full factorisation, but it does not. Terms proportional to
% 1/N vanish at the pole because of the Ward Identity for the sub-amplitudes."} 
% \AL{Suggestion: Switch the order of the two discussions such that it's obvious our main aim is just to explore phase space.
% Then adjust language in soft/collinear from e.g. good to know, to e.g. good to double check the understanding from Mangano/Parke ref that soft/collinear is good}

~While \figsrefd{fig:all-gluon accuracy}{fig:ttbar accuracy}
show the average accuracy of the expansion in a flat phase-space scan,
it is also good to know if the accuracy and precision are dependent on the phase-space region.
In order to check this, 
we looked at the processes $u\ubar \rightarrow 3g$ and $u\ubar \rightarrow 4g$
for $10^7$ phase-space points produced by RAMBO.
For each point, we calculated the energy fractions $x_i = 2E_i/E_{cm}$
of each particle, storing the minimum one;
and calculated the cosine of the opening angle between each particle,
$\cos(\theta_{ij})$, storing the maximum value (minimum angle) for each point.

\begin{figure*}
\resizebox{1\textwidth}{!}{%
  \includegraphics{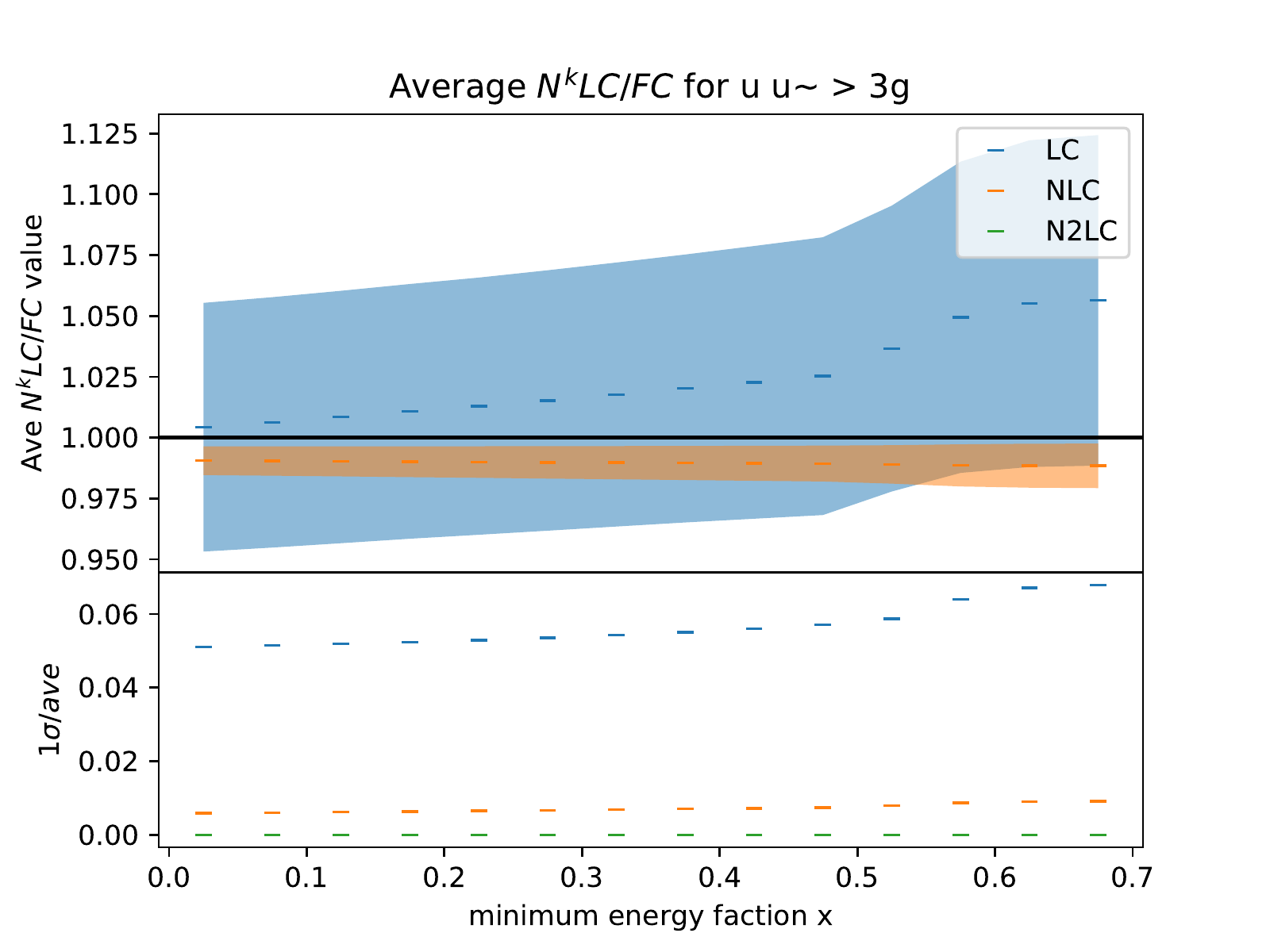}\hfill
  \includegraphics{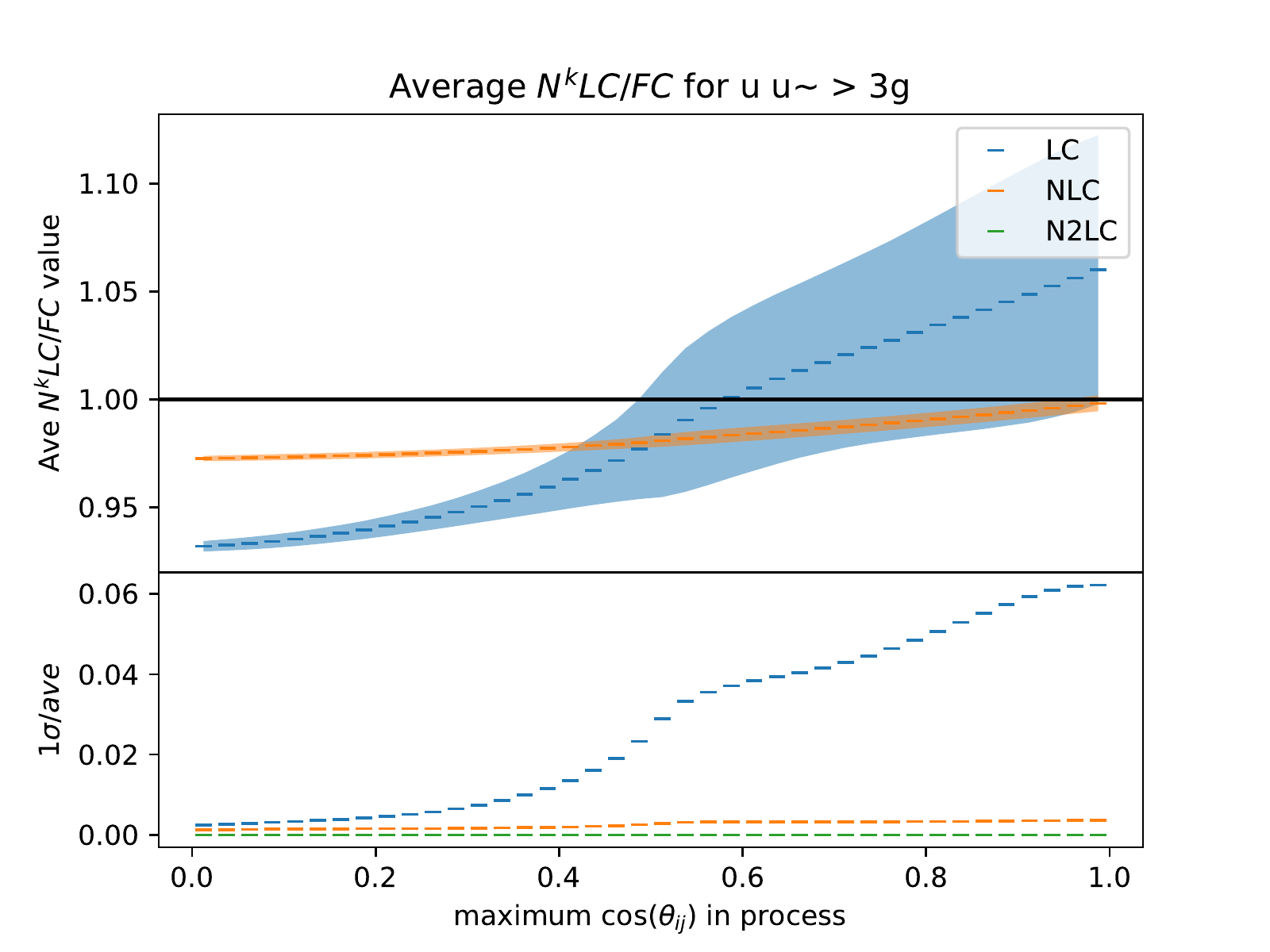}
}\caption{Accuracy and precision of expansion in colours $1/\Nc$ in the process 
$u\ubar \rightarrow 3g$
as a function of the minimum energy fraction $x = \min(2E_i/E_{cm})$
of a given particle (left) and maximum $\cos(\theta_{ij})$ 
between two particles (right).}
\label{fig:soft collinear accuracy 3g}
\end{figure*}

As is shown in \figsrefa{fig:soft collinear accuracy 3g}{fig:soft collinear accuracy 4g},
the accuracy and precision, especially at LC, depends strongly on whether all particles are well-separated or not in angle.
On the other hand, the energy of the softest particle appears to have little importance on the accuracy of the colour expansion. 
Since the accuracy and precision of LC appears to depend too much on the phase-space point,
we think that LC is too crude to be used.
On the other hand, NLC can be used, 
but may vary slightly with the opening angle of two particles,
which might create an issue depending on the multiplicity and how the approximation is used.

\begin{figure*}
\resizebox{1\textwidth}{!}{%
  \includegraphics{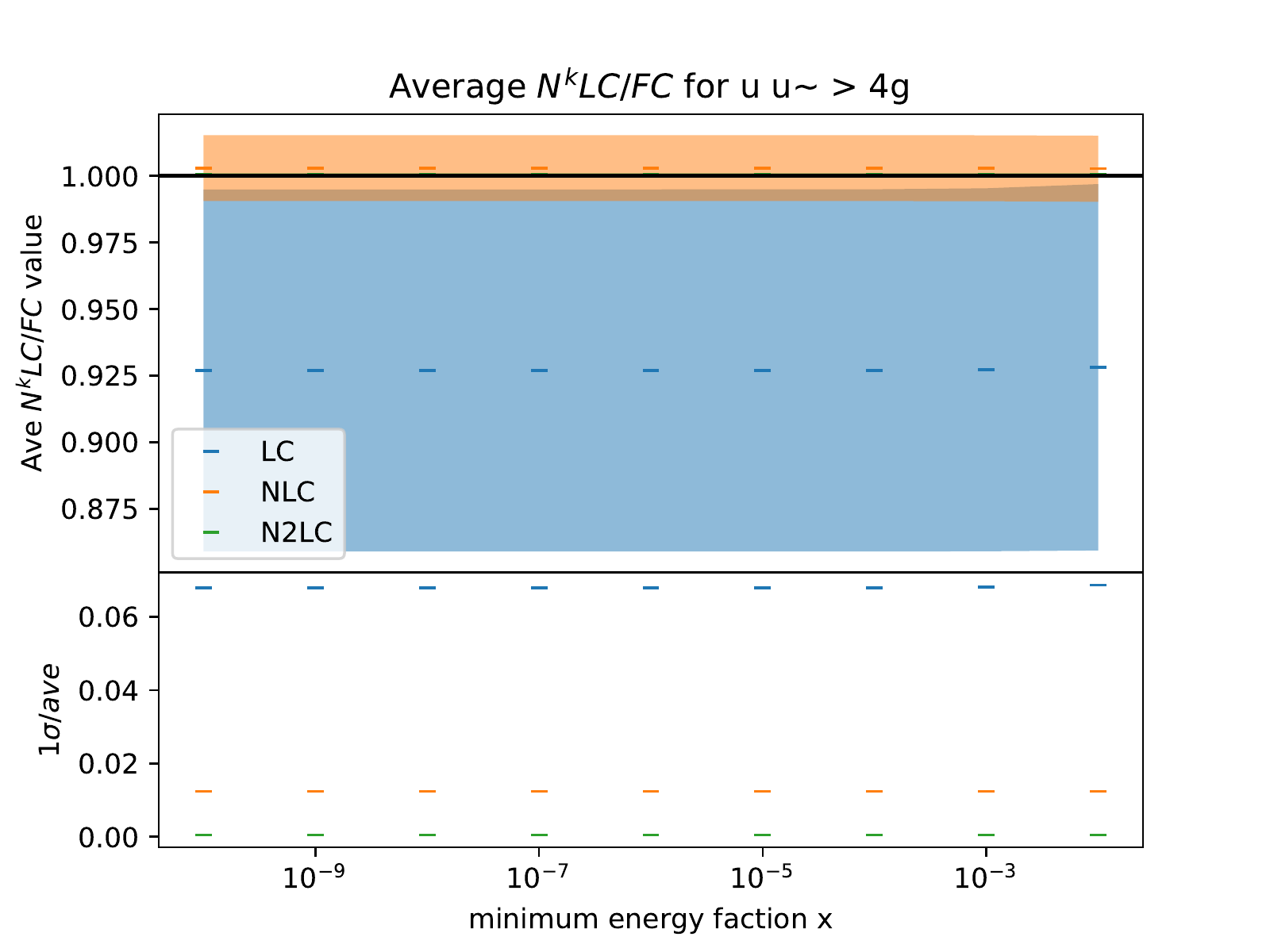}\hfill
  \includegraphics{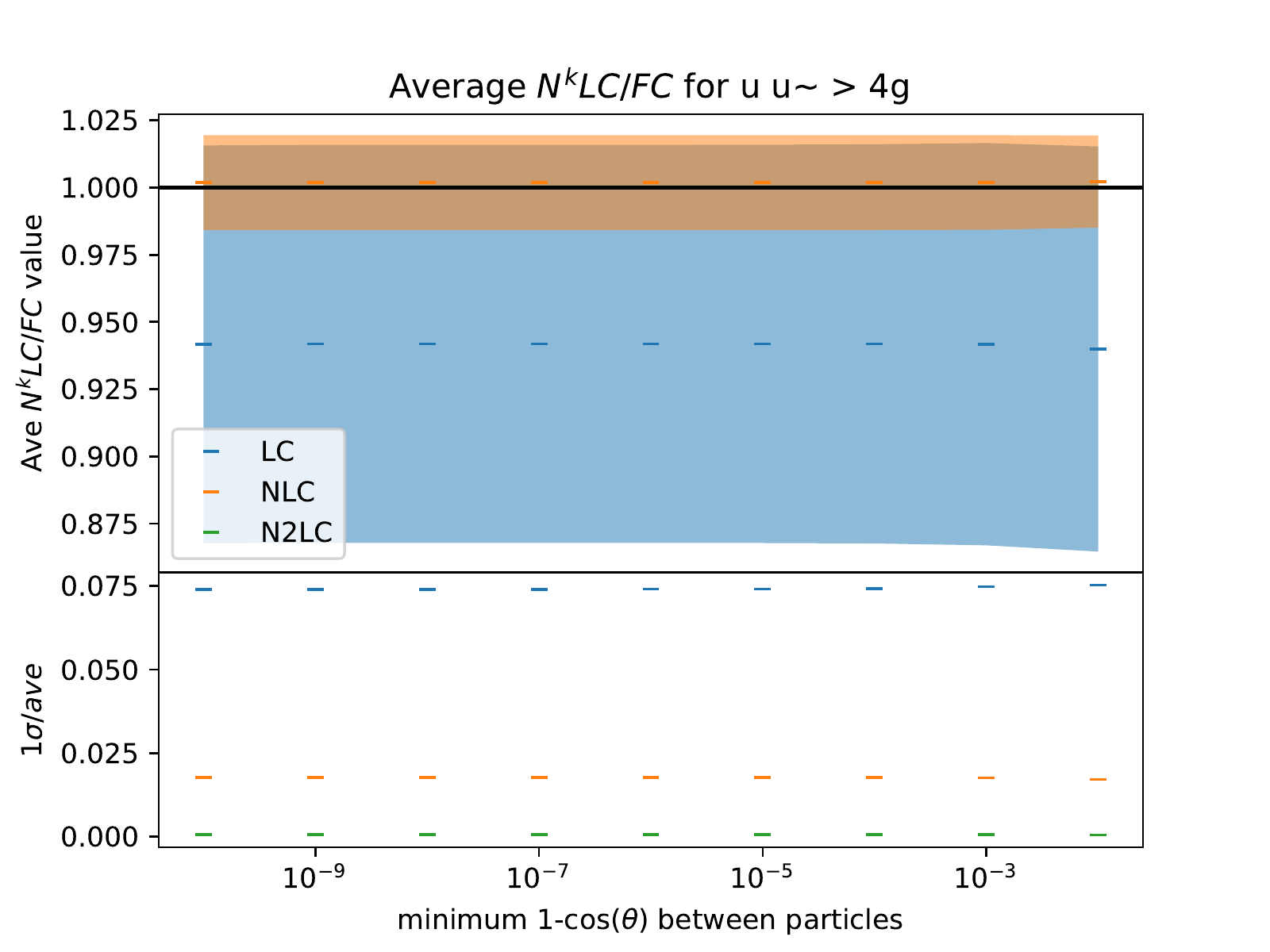}
}\caption{The accuracy and precision of the colour expansion in the soft and collinear limits for
$u\ubar \rightarrow 4g$.}
\label{fig:soft collinear accuracy 4g deep}
\end{figure*}

In addition to this general scan over phase-space, 
it is useful to confirm that each of the colour-ordered amplitudes has the expected soft and collinear limit \cite{Mangano:1987xk}.
To do this, we created around a thousand $u \ubar \rightarrow 3g$ Born phase-space points,
and added a fourth soft or collinear gluon. 
The added gluon was then made more and more soft or collinear to another parton. 
As we see in \figref{fig:soft collinear accuracy 4g deep},
the accuracy and precision of the colour expansion are not changed in the 
deeply soft or collinear limits.
Therefore, the inclusion of the pole in the squared matrix element does not depend on the terms included in the colour expansion.

\begin{figure*}
\resizebox{1\textwidth}{!}{%
  \includegraphics{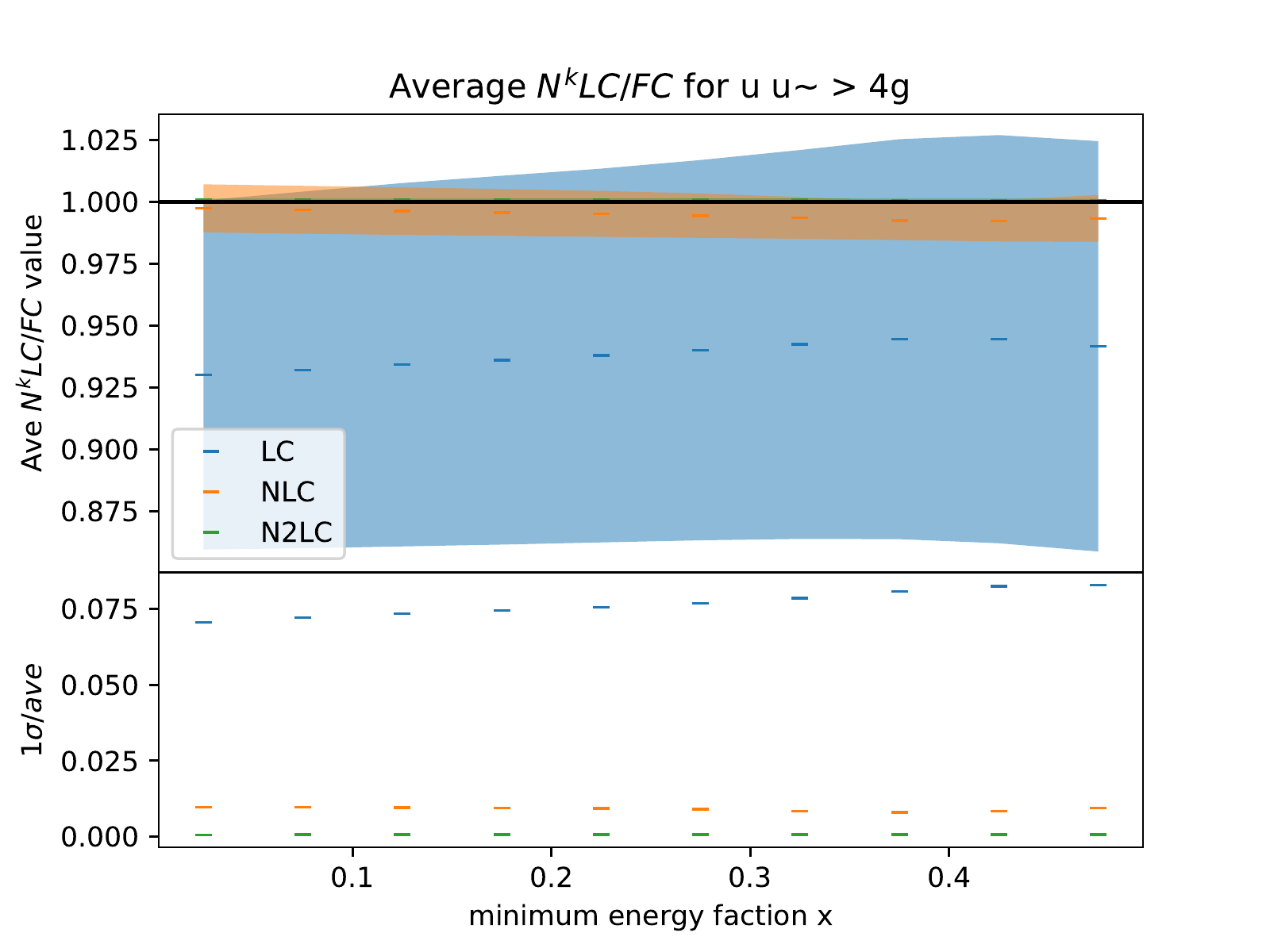}\hfill
  \includegraphics{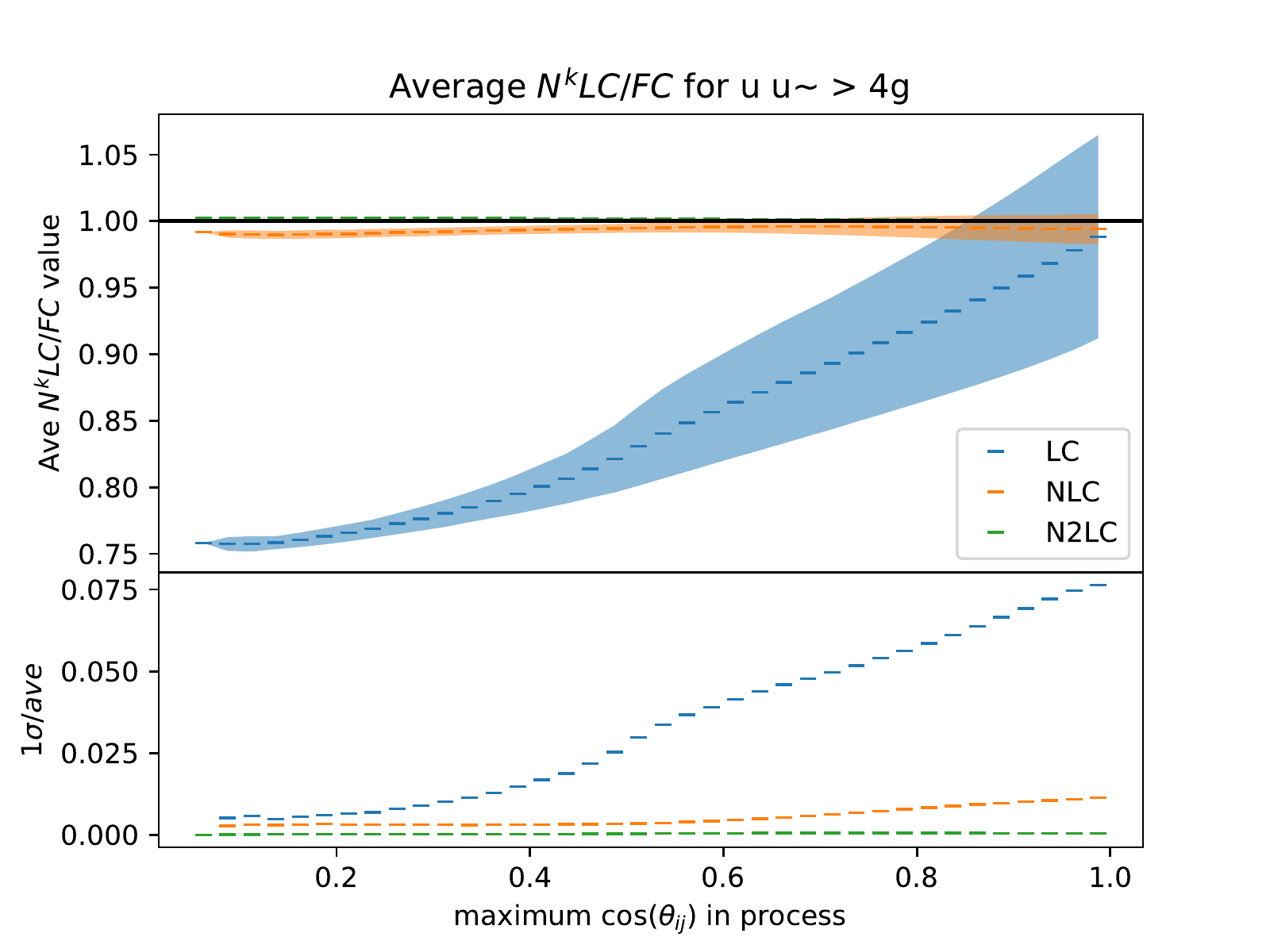}
}\caption{Same as \figref{fig:soft collinear accuracy 3g} but for the process 
$u\ubar \rightarrow 4g$.}
\label{fig:soft collinear accuracy 4g}
\end{figure*}

\subsection{Speed Gain}
\label{sec:speed gain}
% \begin{itemize}
%         \item Do a deep dive into the all-gluon results and comment on accuracy/precision/modLC
%         \item Show other 3 massless QCD procs in single figure and compare results to all gluon
%         \item Show other results here or in the appendix (e.g.\ $t\bar{t},w,z$ + QCD
%         \item Mention that all-gluon interactions were preferred for optimisation
%         \item Mention that generation time is now far improved allowing to go a particle further
%         \item Mention that also all-gluon procs are much more likely in a cross section and take up more time, so not too bad that e.g.\ multiquark amps not optimised
%         \item Mention that we have many options still to optimise further
%         \item Mention that there is nothing stopping us from using the best of both worlds, i.e.\ FC when better and (N)LC etc. when better?
% \end{itemize}

In this section we compare the speed of this new code with that of standard \mgs{}.
To do this,
we compare the time taken to evaluate the same matrix elements in the new and old codes
(for the different sources of speed gain (and loss),
see \secref{sec:sources of speed differences}).
Note that these comparisons ignore the time taken to generate and compile the code in the new and old way\footnote{
The new code is generated and compiled much faster at high multiplicity,
and both codes take a similar time to generate and compile at low multiplicity.}.

\begin{figure}
\resizebox{0.5\textwidth}{!}{%
  \includegraphics{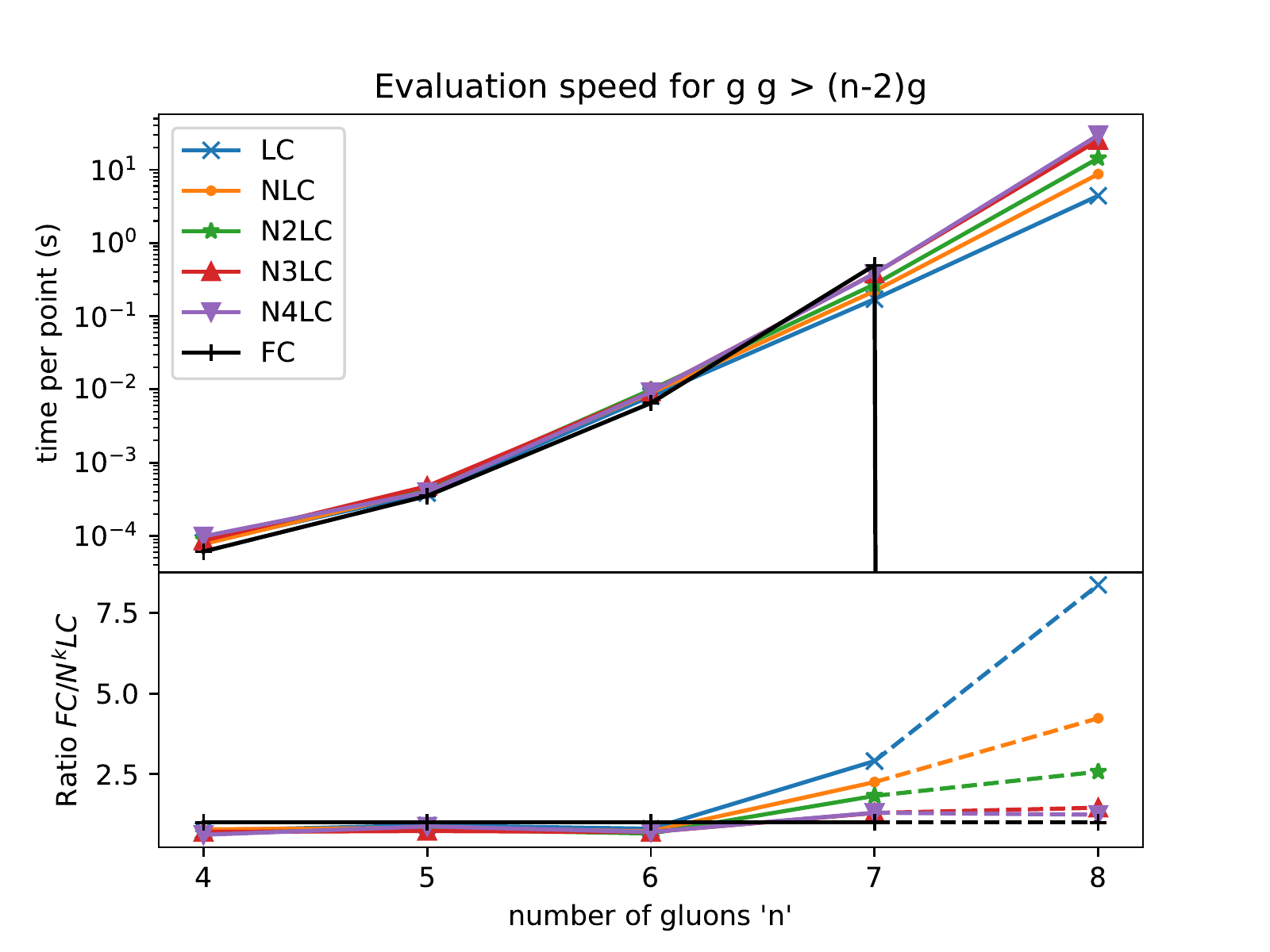}
}
\caption{
Top: speed of the process $gg \rightarrow (n-2)g$ 
for each colour ordering and gluon multiplicity.
FC corresponds to standard \mgs{} (version 2.9.2, for standalone essentially equivalent to the latest version 3.4.1).
Bottom: Ratio of the speed using standard \mgs{} to using the new code for each colour ordering. 
Standard \mgs{} cannot calculate $gg \rightarrow 6g$,
so the right-most ratio speeds instead contains the N5LC BG speed on the denominator to show the effects of colour ordering and obtain an estimate for the true speed increase. 
Speed tests done on a MacBook Pro 2020 CPU i5-8257U.}
\label{fig:all-gluon speed}       % Give a unique label
\end{figure}

\paragraph*{All-gluon amplitudes:}
~First we describe in detail the speed of the all-gluon amplitudes,
shown in \figref{fig:all-gluon speed}.
The top panel of this figure shows the average time it takes to calculate a single phase-space point at each gluon multiplicity and each order of the colour expansion.
The bottom panel shows the ratio
\begin{equation*}
    \frac{t_{FC}}{t^{new\, code}_{N^kLC}},
%    \frac{\text{time taken using standard \mgs{}}}{\text{time taken using BG with colour at given order in $1/\Nc$ }}~,
\end{equation*}
% AL 20220915: checked that fraction corresponds to plots
where $t_{FC}$ is the time taken using standard \mgs{},
and $t^{new\, code}_{N^kLC}$ is the time using the new code (with BG recursions) with the colour matrix expanded to include all terms up to $N^k$LC.
It allows to quantify the speed gain or loss from using the new code and truncating the colour expansion. 
When the order in $1/\Nc$ is high enough, both the old and new codes are evaluating the same matrix element and there is no speed gain due to truncating the expansion.

At low gluon multiplicity, 
the new code is actually slower than standard \mgs{},
but at seven gluons the colour sum dominates sufficiently such that the new code is  between 1.2 and 2.9 times faster than 
\mgs{} depending on the truncation of the colour expansion,
and at eight gluons, we can only use the new code.
We therefore significantly speed up the slowest processes,
even though we slow down some faster ones.
% (recall that many possible optimisations remain which should hopefully fix this problem).

There are  several options to address the speed loss.
The first is to optimise the BG recursions.
As discussed in \secref{sec:BG recursion implementation}, there
 are many possible optimisations not yet used in the BG recursion, 
and implementing them should help alleviate this problem.
A second option is to import the colour computation from the new code into standard \mgs{} and ignore BG recursions completely.
A third option is to use some optimised BG recursions and the new colour computation at high multiplicity,
and use standard \mgs{} together with the new colour computation at low multiplicity.
Since BG recursions are expected to bring gains at high multiplicity,
this may create a best of both worlds scenario. 
Exploring these options is left for future work.

Since we cannot use standard \mgs{} for 8 gluons,
the speed increase for this process is compared to the N5LC BG recursion in the ratio plot at the bottom of \figref{fig:all-gluon speed},
i.e.\ the increase shown is purely due to truncating the colour matrix.
This is almost certainly an underestimate of the speed increase.

It is worth noting that since the colour matrix has size $(n-1)!\times (n-1)!$,
the effect of truncating the matrix leads to larger speed gain for larger gluon multiplicity.
By 8 gluons, the LC amplitude is over 8 times faster than the full answer calculated with BG recursions,
while the N2LC result is over twice as fast
(recall that the 8 gluon N2LC amplitude is accurate to within a few percent and has a precision of about half a percent, 
see \figref{fig:all-gluon accuracy}). 
At 7 gluons the LC result is about 2.4 times faster than the FC result when FC is calculated using the new code (i.e.\ when the only difference is the truncated colour matrix).

\begin{figure}
\resizebox{0.5\textwidth}{!}{%
  \includegraphics{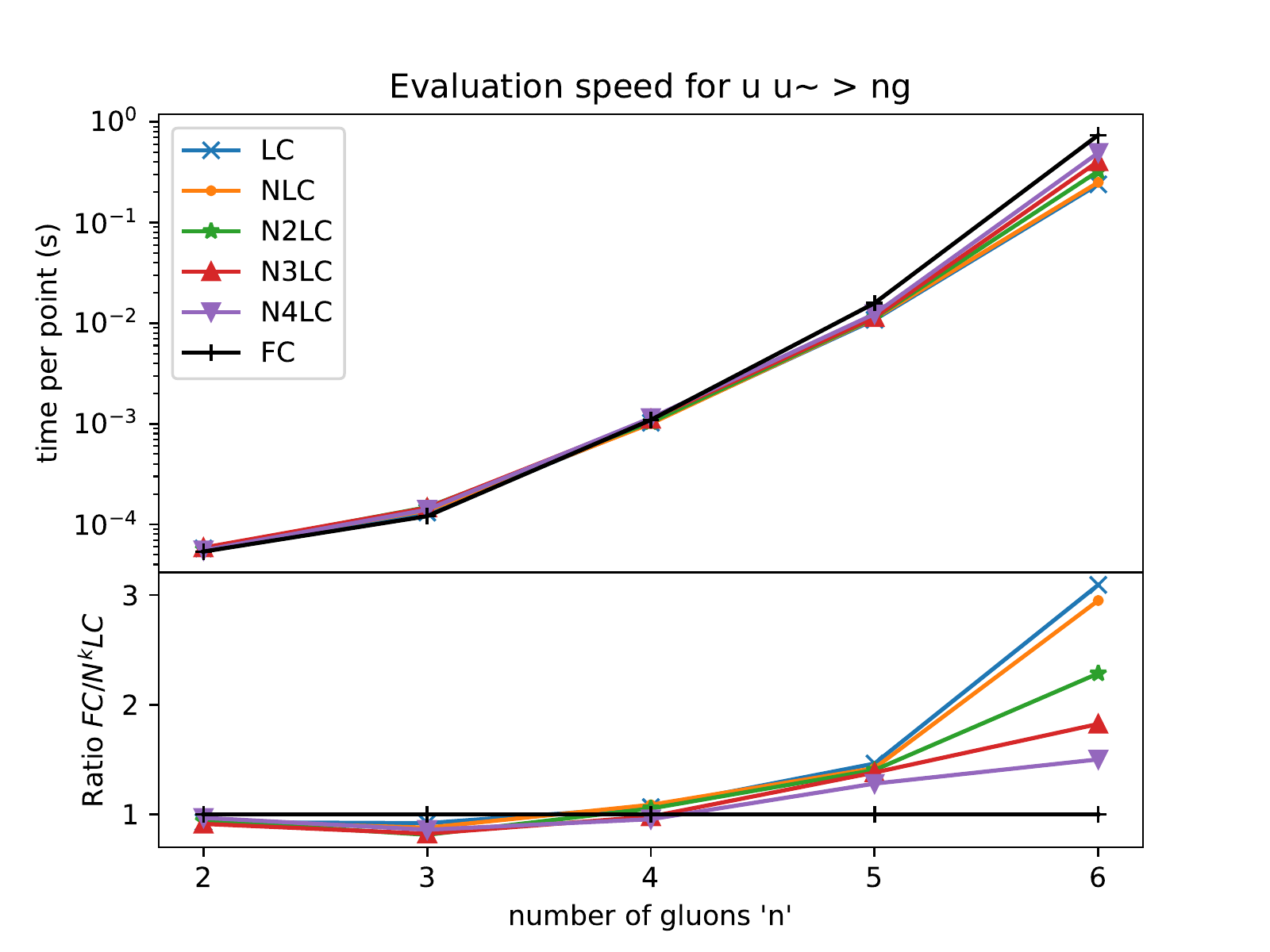}
}
\caption{Same as \figref{fig:all-gluon speed} but for $u\ubar \rightarrow ng$.}
\label{fig:quark speed}       % Give a unique label
\end{figure}

\paragraph*{Amplitudes with a single quark pair:}
~Next, we consider QCD processes with a single quark pair, 
again using $u\ubar \rightarrow ng$ as a test process (see \figref{fig:quark speed}).
We again see that the new code is much faster at high gluon multiplicity,
and a bit slower at low gluon multiplicity. 
This amplitude is about a factor 10 faster than the all-gluon amplitude,
and has a similar level of importance
(see \appref{mlm}, \figref{fig:cross-section}).
The 6g amplitude at N2LC is about 2.3 times faster than standard \mgs{} 
with an accuracy of around $0.1\%$ and precision of around $0.5\%$
(see \figref{fig:quark accuracy}).

\begin{figure*}
\resizebox{1\textwidth}{!}{%
  \includegraphics{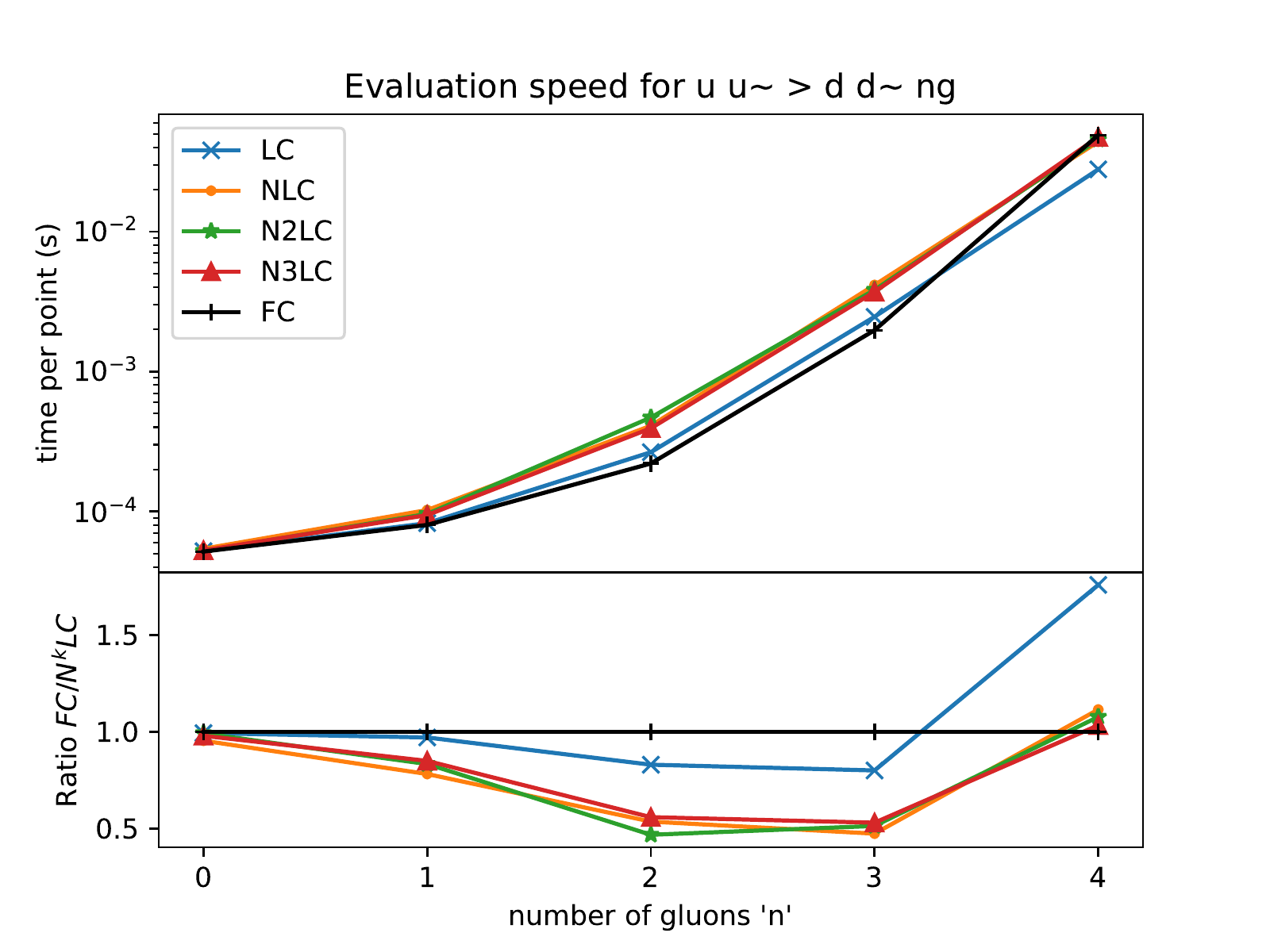}\hfill
  \includegraphics{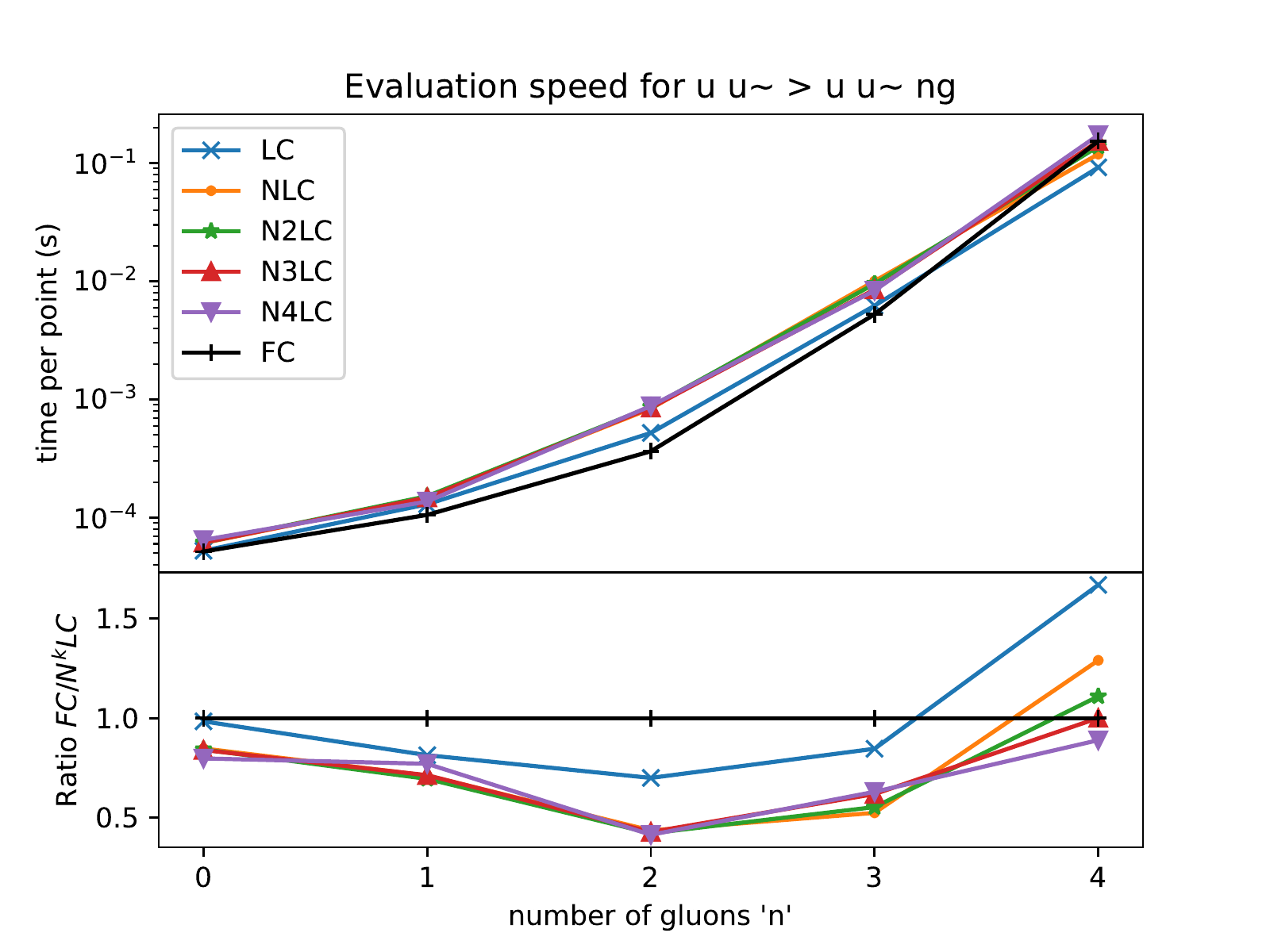}
}\caption{Same as \figref{fig:all-gluon speed} but for $u\ubar \rightarrow d\dbar + ng$ (left) and $u\ubar \rightarrow u\ubar + ng$ (right).
}
\label{fig:multiquark speed}
\end{figure*}

\paragraph*{Amplitudes with two quark pairs:}
~To complete the pure massless QCD analysis,
we again study $u\ubar \rightarrow d\dbar + ng$
and $u\ubar \rightarrow u\ubar + ng$ as shown in \figref{fig:multiquark speed}.
This time the new code is significantly slower than standard \mgs{} for low gluon multiplicity, 
but again starts to become faster at high multiplicity.
%This is because the new program was only optimised for permutations of gluons, 
%and still requires optimisation for permutations of quarks. 
%\AL{also multiple flows?}.
However, as one can seen in \appref{mlm} (\figref{fig:cross-section}),
this process is less significant than the other massless QCD processes.
Further, comparing \figref{fig:multiquark speed} to
\figsrefa{fig:all-gluon speed}{fig:quark speed},
we see that multiquark amplitudes are also quicker than most other massless QCD processes,
hence a speed gain or loss here is not so significant.

\begin{figure*}
\resizebox{1\textwidth}{!}{%
  \includegraphics{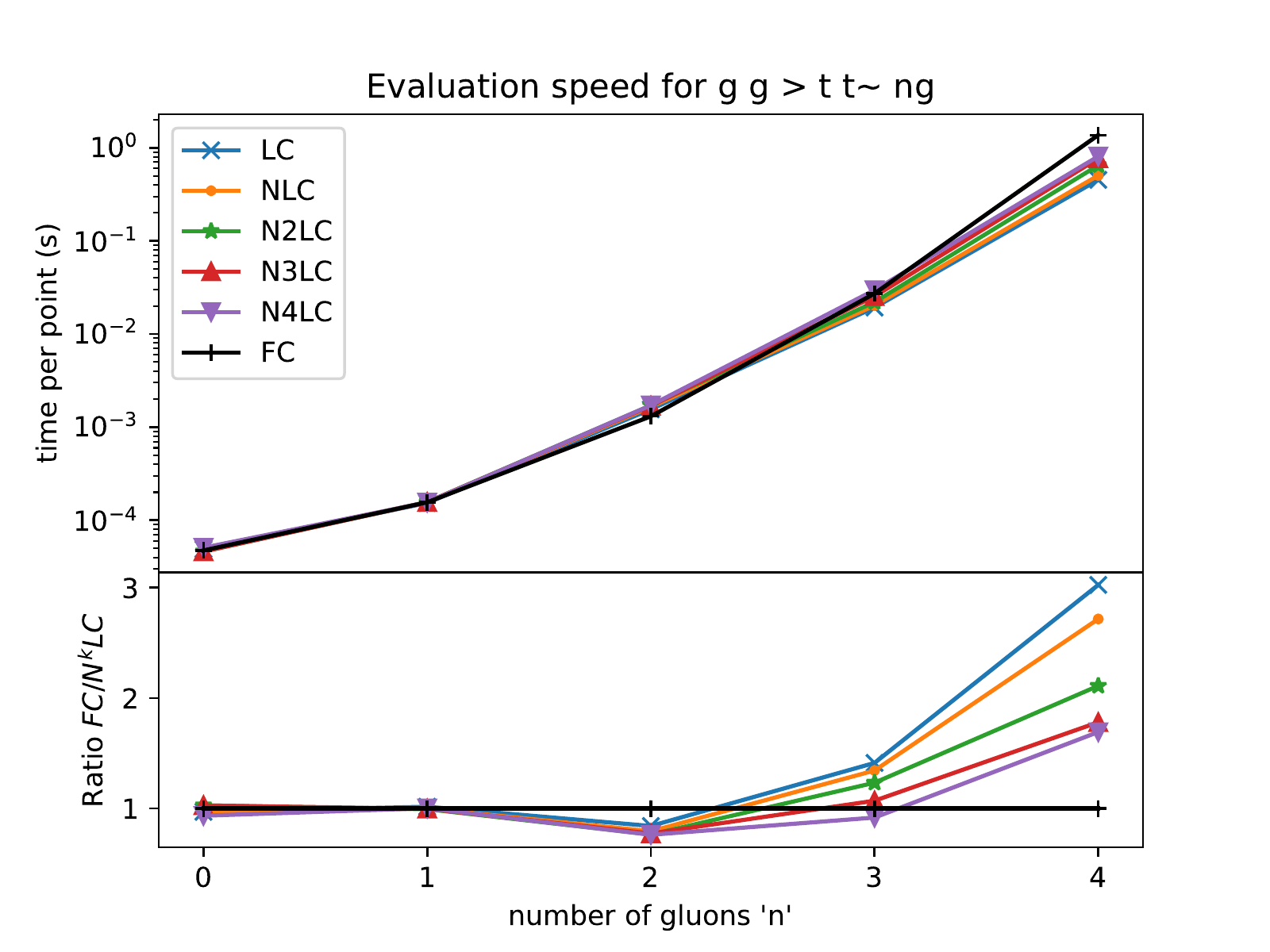}\hfill
  \includegraphics{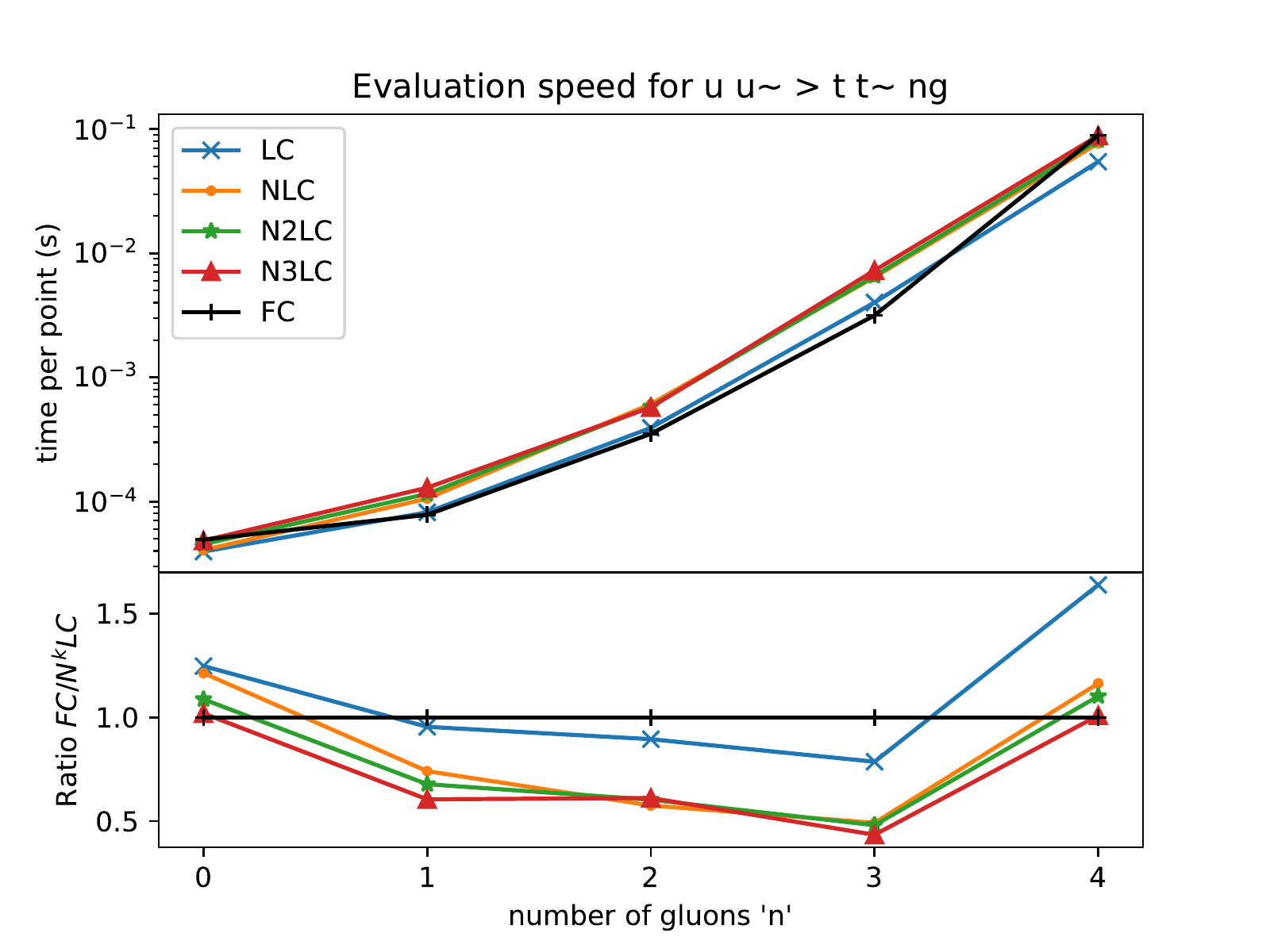}
}\caption{Same as \figref{fig:all-gluon speed} but for $gg \rightarrow t\tbar + ng$
and $u\ubar \rightarrow t\tbar + ng$.}
\label{fig:ttbar speed}
\end{figure*}

\paragraph*{Amplitudes with a top quark pair:}
~Finally, in \figref{fig:ttbar speed},
we consider the speed of pure QCD processes with a top pair.
% (\figref{fig:ttbar speed}). 
Once again, at high multiplicity (in this case four or more gluons in the final state)
we see the new code becomes faster than standard \mgs{}.
For less final-state gluons the old code is quicker. 
% \OM{add a reference to the figure.}

% calculating the wavefunctions, both external and internal (i.e.\ propagators or off-shell BG currents);
% calculating the amplitudes, i.e.\ completed Feynman or BG graphs;

\section{Conclusion}
\label{sec:conclusion}
% \begin{itemize}
%     \item Conclude that we have implemented both BG recursion and a $1/\Nc$ expansion in \mg.
%     \item Summarise some of the key results
%     \begin{itemize}
%         \item We can calculate amps at NLC at quite good precision so can likely correct 
%         \item LC is not a good target for accurate or precise predictions at high multiplicity
%         \item We can do some amplitudes within \mgs{} that we couldn't before (even though was possible in other codes)
%         \item For high multiplicity, even at full colour, the BG recursions offer a faster calculation than current \mgs{} and has given us insight into how to optimise the colour calculations also in normal \mgs{} 
%         \item We still have factorial growth in colour matrix
%     \end{itemize}
%     \item future work
%     \begin{itemize}
%         \item Moving to madevent (possibly via Timea + Rikkert) (MC over colours for high colours?)
%         \item Getting the colour factors in a smarter way (i.e.\ work of Timea + Rikkert)
%         \item Optimise the kinematics of the code (might slow generation considerably)
%         \item Optimise colour computations in main \mgs{} code
%         \item Use approximation trick from NLO for NLC amp?
%     \end{itemize}
% \end{itemize}

In this paper, we have re-implemented the colour computation of \mgs{}
and implemented BG-like recursions within \mgs{}. 
We now have both a more efficient way to generate QCD amplitudes,
as well as a faster matrix-element evaluation at high multiplicity. 
In particular, \mgs{} 
can for the first time generate and evaluate matrix elements for
$g g \rightarrow 6g$ and some other high multiplicity processes.

For the colour computation, 
we defined an expansion of the colour-matrix as a function of the highest power of $\Nc$, 
and studied the accuracy and relative precision of the expansion for various processes. 
In general the LC approximation does not provide either an accurate or precise value of the full matrix-element squared,
and therefore is barely usable for any practical application. 
The situation radically improves for NLC accuracy where the precision is typically at the percent level,
even if the computation can be affected by a large bias.
This approximation should be enough to speed up phase-space integration, 
thanks to various phase-space integration methods based on having access to fast matrix-elements \cite{Alwall:2014hca,Danziger:2021eeg,Weinzierl:2000wd}. 
For the all-gluon amplitude, the N2LC approximation is also affected by a bias. 
However, all other processes are precise at the per-mil level at N2LC and do not have any significant bias. 
In all cases, the N3LC amplitudes are extremely precise and accurate,
and should be usable without corrections in many applications.

Importantly, the novel implementation of the colour sum in the new code improves the evaluation time of high-multiplicity matrix elements,
even without truncating the colour expansion.
If truncating the colour expansion,
we can further gain in the evaluation time by using phase-space symmetry to limit the number of colour orderings required \cite{Frederix:2021wdv}.
% \AL{Add comment on new way to calculate even full colour matrix making things faster?}. 
At low multiplicity, 
the computation of the colour-matrix is not critical,
and since our implementation of the BG relation is not as optimised as standard \mgs{}, the new code is slower than the old code at these multiplicities. 
Such optimisation is left for future work. 
Additionally, like done in \cite{Frederix:2021wdv},
it would be beneficial to know in advance which terms of the colour matrix contribute to which order of the expansion.
This would greatly help speed up the generation of the code, allowing to go to even higher multiplicity.

This paper is an important milestone for the \mgs{} code, 
both by allowing higher multiplicity,
and by allowing more control on the colour treatment of the computation.
Now such improvement needs to be incorporated within the other types of computation offered by \mgs{},
in particular for LO/NLO cross-section/event generation for merged generation.
The best approach here would require some deep change within the phase-space integrator since it is not compatible with BG recursions \cite{Maltoni:2002qb}.
Independently of making these deep changes, importing the new colour computation into the main code should be fairly straightforward.
This optimisation should allow to have, for high multiplicity, 
code faster by around thirty percent, 
thus allowing us to meet the requirement needed for HL-LHC \cite{HEPSoftwareFoundation:2020daq,Collaboration:2802918}.

\section*{Acknowledgements}
We would like to thank Rikkert Frederix, Malin Sjödahl, and  Timea Vitos for a thorough reading of, and comments on, the manuscript. We would also like to thank Johan Alwall, Stefano Frixione, and Fabio Maltoni for discussions related to this paper. AL would like to additionally thank Malin Sjödahl for the encouragement to branch out during his PhD and do this project.

This work has received funding from the European Union's Horizon 2020 research and innovation programme as part of the Marie Skłodowska-Curie Innovative Training Network MCnetITN3 (grant agreement no. 722104).
In addition, AL would like to thank his funding from the Swedish Research Council (contract number
2016-05996, % VR group contract by Torbjörn
as well as European Union's Horizon 2020 research and
innovation programme (grant agreement No 668679). % Torbjörns group contract from European Union
OM received funding from FRS-FNRS agency via the IISN maxlhc convention (4.4503.16).
Computational resources have been provided by the supercomputing facilities of the Université catholique de Louvain (CISM/UCL) and the Consortium des Équipements de Calcul Intensif en Fédération Wallonie Bruxelles (CÉCI) funded by the Fond de la Recherche Scientifique de Belgique (F.R.S.-FNRS) under convention 2.5020.11 and by the Walloon Region.

\appendix

\section{Manual}
\label{sec:manual}

In this section, we describe how to use the new code.
This short manual will assume the reader is familiar with and already knows how to run \mgs{}
(the unfamiliar reader is directed to \cite{Alwall:2014hca} for the structure and main commands of \mgs{}).
We remind that the recursions and the colour ordering are only available in standalone mode,
such that we can only calculate squared matrix elements, and not cross sections 
(which would require a dedicated phase-space integrator).
Both the BG recursions and new colour implementation can only be used within a colour expansion. 
If the user wants the full colour amplitude using the new code,
they simply need to choose a high enough colour order such that the expansion finishes.
For example, if the expansion naturally terminates at NLC, 
then choosing colour order NLC, N2LC, N3LC etc. will all give the full colour result.
There is no known time penalty for choosing e.g. N3LC when the expansion naturally terminates at NLC.

To switch on the new code we use the \tet{set} \tet{color\_order\-ing} command, 
where \co{} 0 means using normal \mgs{} and is the default.
If \co{} is set to $k\geq 1$, then \mgs{} will calculate the $\text{N}^{k-1}\text{LC}$
amplitude.
Only \textit{after} setting the colour ordering to a non-zero value is it possible to use 
\tet{set optimization} to toggle between BG recursion (default, 
\tet{optimization 3}) and standard Feynman diagrams (\tet{optimization 1}).
Note that although \tet{optimization 1} uses Feynman diagrams,
it does \textit{not} use the optimised version from standard \mgs{}.
Therefore, using \tet{optimization 1} will be slower than using the default BG recursions.
Finally, if we want to use the modified definition of multiquark colour (see \appref{sec:modLC multiquark}), 
we can change the \tet{LC\_defn} from its default value of \tet{fund} to \tet{modLC}.

To make the instructions more explicit, 
we write here a sample card (assumed to be called \tet{example.txt}),
which will instruct \mgs{} to generate and calculate the process $pp \rightarrow 5j$
at NLC using BG recursion for the kinematics.
To use it, type  \tet{./bin/mg5aMC example.txt}
in the \mgs{} directory.

\begin{lstlisting}[language=Bash]
 set ignore_six_quark_processes j # avoid 6 quark process
 set color_ordering 2
 set LC_defn fund #default: other option modLC
 set optimization 3 #default: other option 1
 generate p p > 5j
 output standalone
 launch
\end{lstlisting}

\section{Accuracy and Speed of Additional Processes}
\label{sec:EW results}
In this section we briefly repeat the analysis of \secref{sec:validation and results}
for QCD processes with three quark pairs and QCD processes with the addition of an electroweak boson.

\begin{figure*}
\resizebox{1\textwidth}{!}{%
  \includegraphics{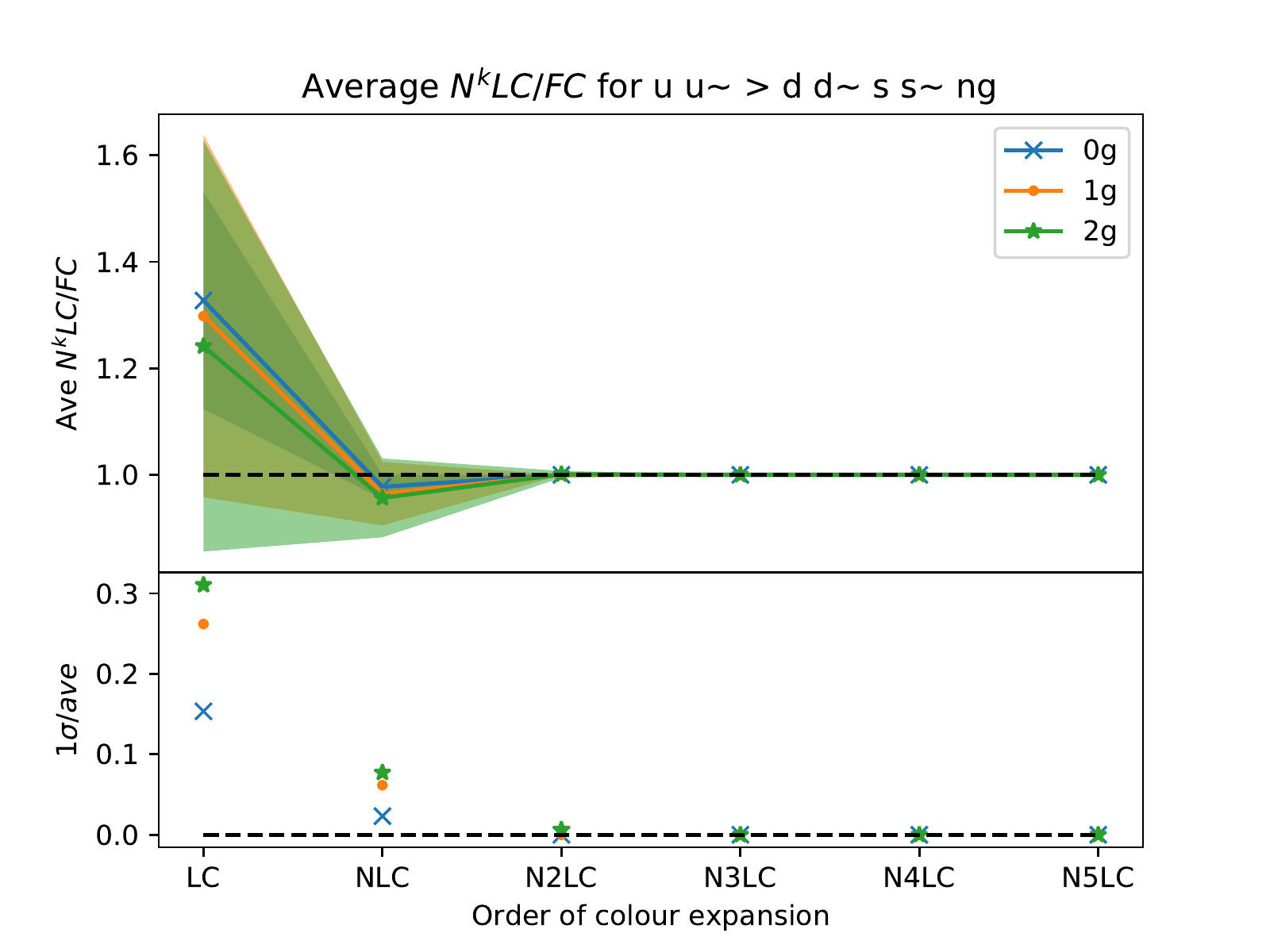}\hfill
  \includegraphics{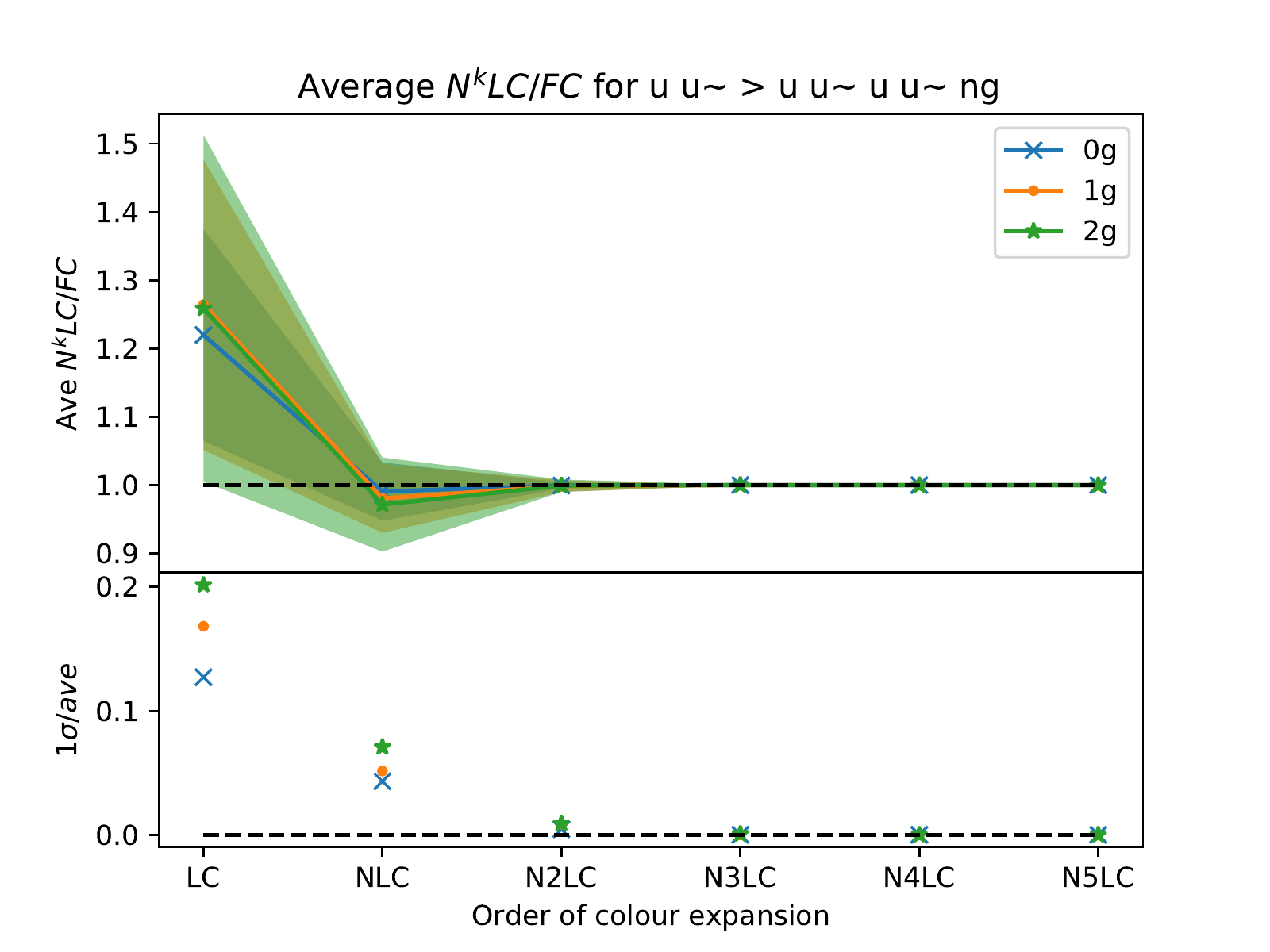}
}\caption{Same as \figref{fig:all-gluon accuracy} but for the processes 
$u\ubar \rightarrow d\dbar s\sbar + ng$
and $u\ubar \rightarrow u\ubar u\ubar + ng$.}
\label{fig:three quark accuracy}
\end{figure*}

\paragraph*{Processes with three quark pairs:}
~Similar to the two-quark amplitudes, we here distinguish between whether the quarks all have the same flavour, 
whether two quark lines have the same flavour, or all quarks are distinct, 
with the first and last of these cases shown in \figref{fig:three quark accuracy}. 
For up to 1 gluon in the final state, we used 100,000 phase-space points. 
For 2 gluons we used 10,000 phase-space points. 
Like in the two-quark-line case (cf \figref{fig:multiquark accuracy}),
the LC accuracy and precision is rather poor,
NLC provides a good approximation, and by N2LC the approximation is very close to exact for the multiplicities studied.

Also, since we optimised for multigluon amplitudes and not for multiquark amplitudes, 
we found the new code to be slower than the old one for this type of process.
This is unlikely to be an issue however, since processes with three quark pairs are typically very sub-leading,
so this matrix element is calculated far less often compared to those in the main text (see \figref{fig:cross-section}).

\begin{figure*}
\resizebox{1\textwidth}{!}{%
  \includegraphics{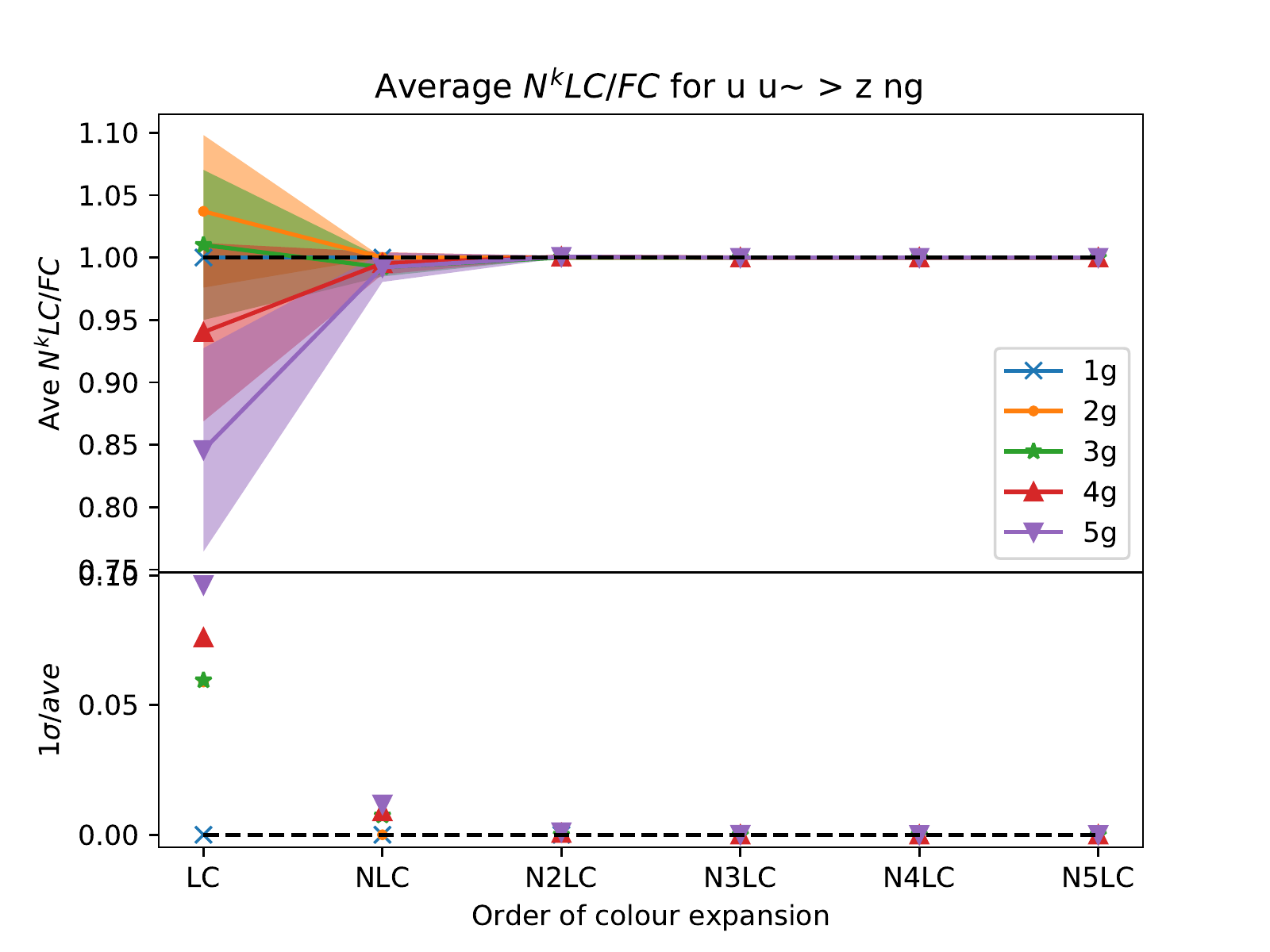}\hfill
  \includegraphics{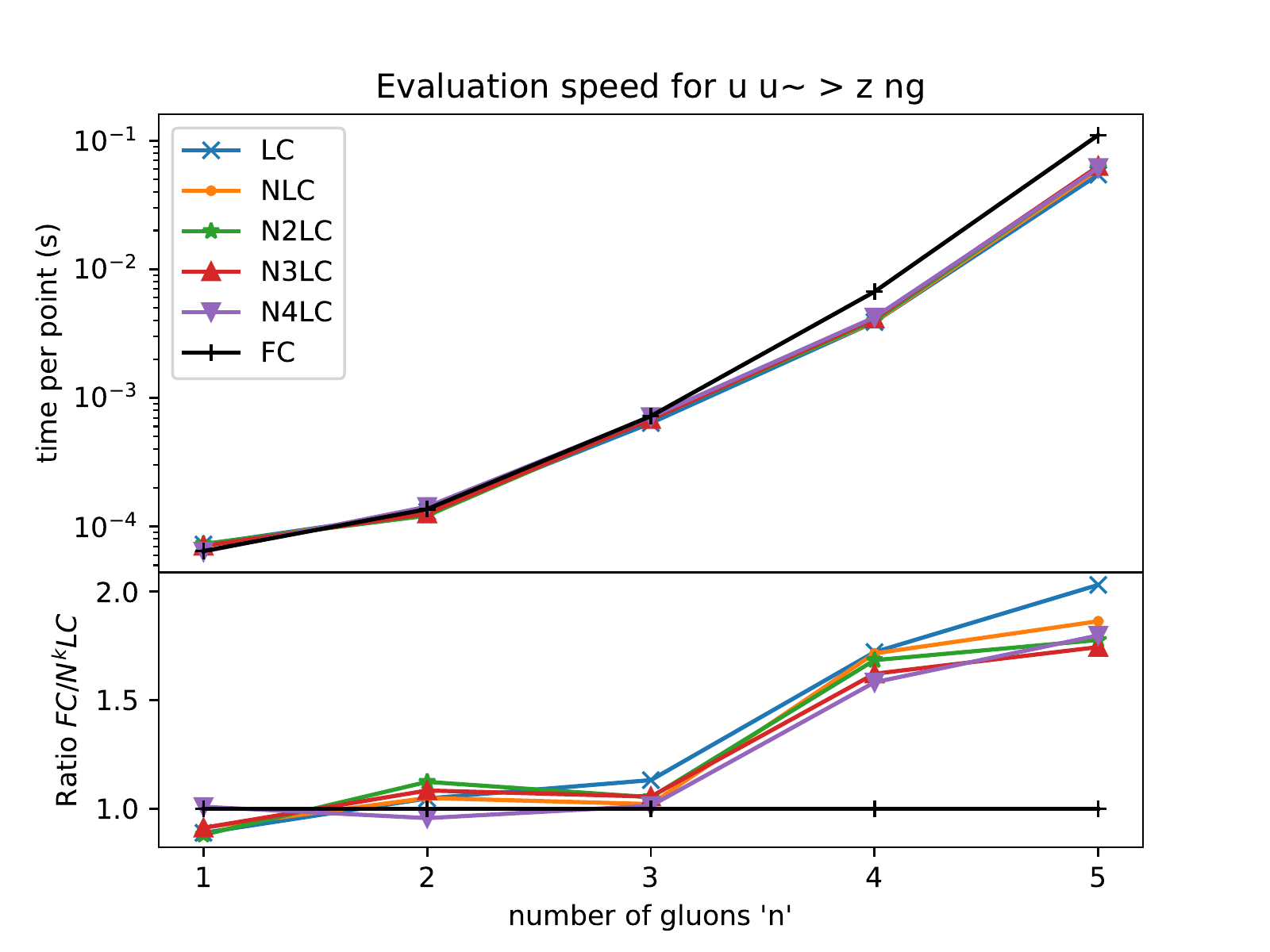}
}\caption{Same as \figsrefa{fig:all-gluon accuracy}{fig:all-gluon speed} 
but for $u\ubar \rightarrow Z + ng$.}
\label{fig:2qz accuracy and speed}
\end{figure*}

\paragraph*{Processes with an EW boson:}
~As a first test case we look at $Z$ production, using the process $u \ubar \rightarrow Z + ng$
(see \figref{fig:2qz accuracy and speed}).
Comparing to \figref{fig:quark accuracy},
we see a similar accuracy and relative precision when the number of gluons are the same.
On the other hand, comparing to \figref{fig:quark speed} and in particular looking at the high multiplicity end,
we see a greater improvement in speed when adding gluons,
but a slightly lower speed gain if comparing overall particle multiplicity.

\begin{figure*}
\resizebox{1\textwidth}{!}{%
\begin{subfigure}{1\textwidth}
  \includegraphics{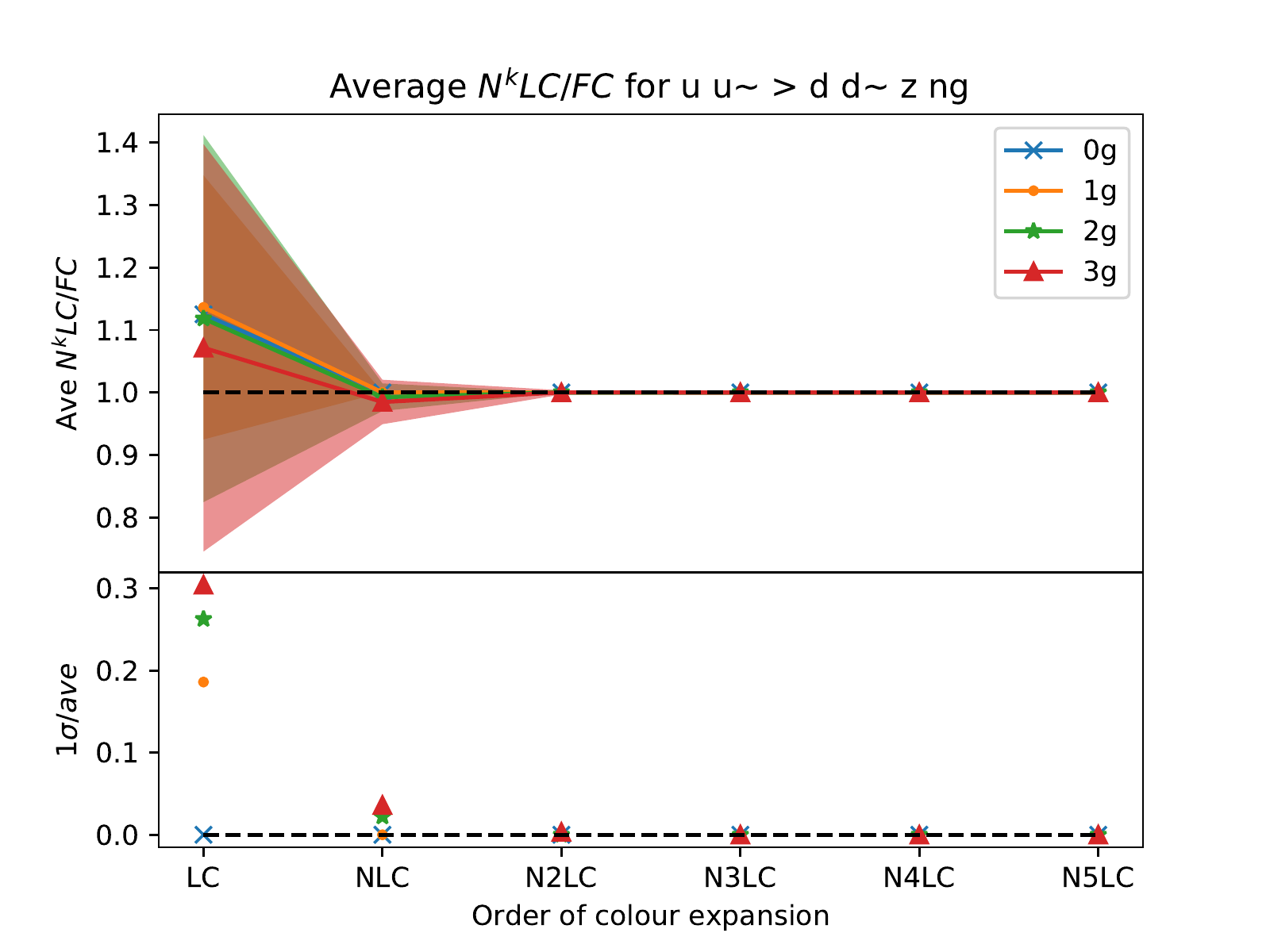}\hfill
  \includegraphics{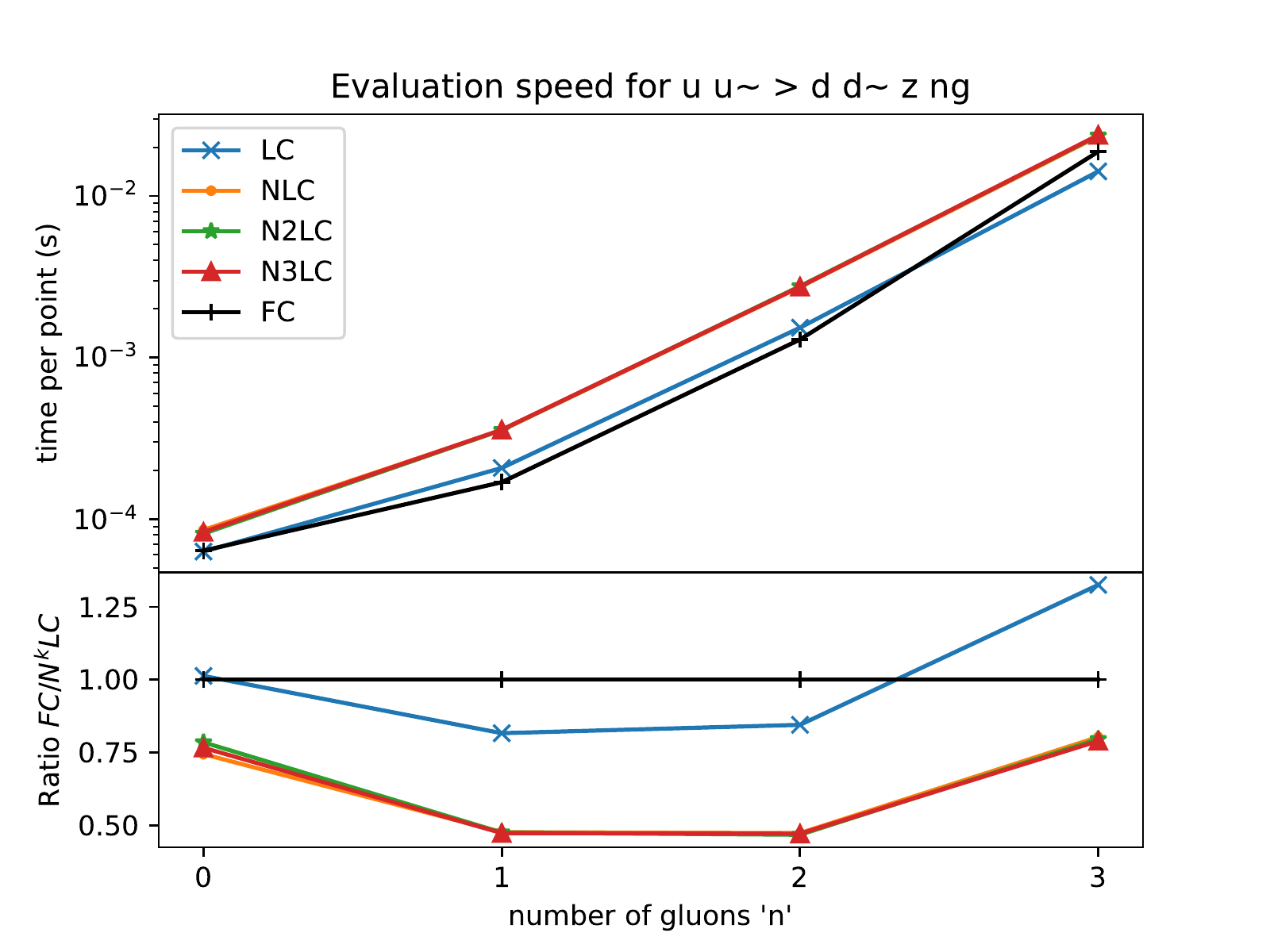} \\
\end{subfigure}
\begin{subfigure}{1\textwidth}
 \includegraphics{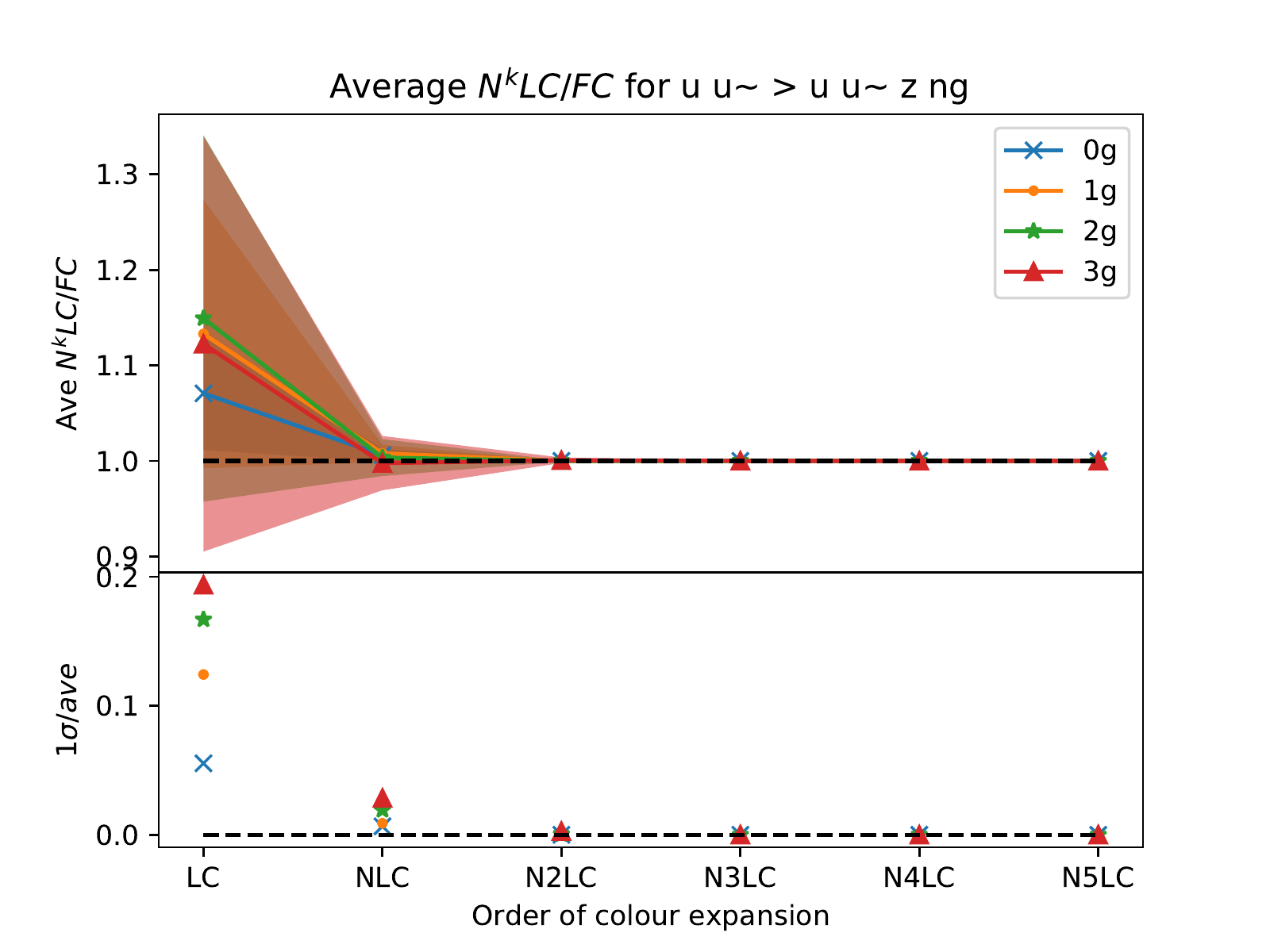}\hfill
  \includegraphics{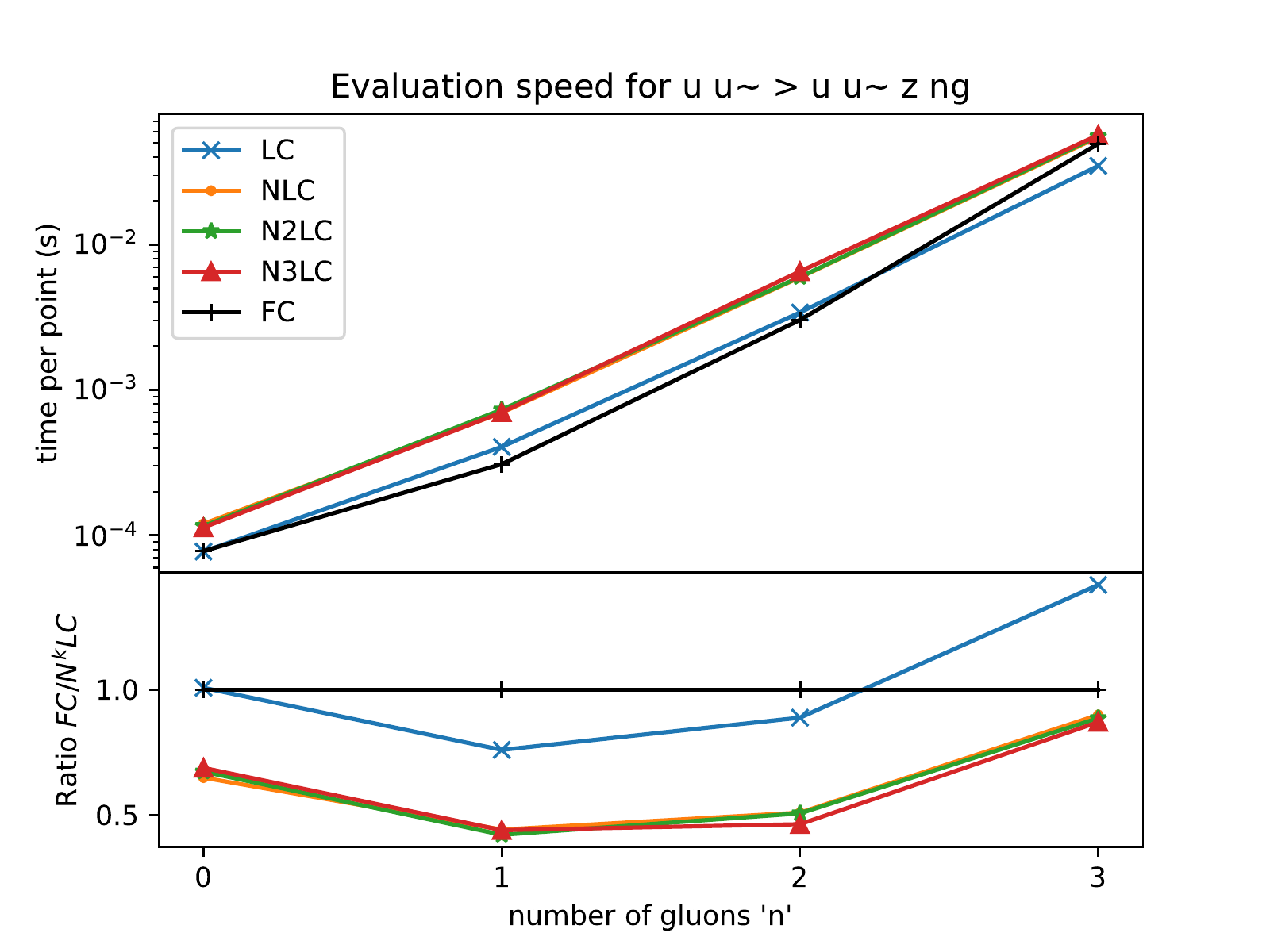}
\end{subfigure}
}\caption{Same as \figsrefa{fig:all-gluon accuracy}{fig:all-gluon speed} 
but for $u\ubar \rightarrow Z \ d\dbar + ng$ (left) and $u\ubar \rightarrow Z \ u\ubar + ng$ (right).}
\label{fig:4qz accuracy and speed}
\end{figure*}

As a second test case we consider $Z$ boson production with an additional quark pair
(see \figref{fig:4qz accuracy and speed}).
Comparing to \figref{fig:multiquark accuracy} 
we again see that the accuracy and precision is largely driven by the number and nature of the QCD particles involved.
Instead comparing the speed to \figref{fig:multiquark speed},
we see that this time we have a worse speed performance when adding a $Z$ 
boson compared to the pure QCD multiquark case.

We conclude that the $Z$ boson has little effect on the accuracy and precision,
and that it is the QCD part of the process which is important for this.
On the other hand, the $Z$ boson has a large role to play in evaluation speed.
% AL 220812: LC diff in accuracy for 2q typically is 0.01, 4qDiff typically order 0.001, 4qSame typically few percent
% AL 220812: LC diff in rel precision for 2q typically is sub percent, 4qDiff typically order percent, 4qSame typically percent or less
% AL 220812: NLC diffs even smaller than LC

Similar to the $Z$ boson, we tested the addition of a $W$ boson by testing the processes
$u \dbar \rightarrow W^+ + ng$, $u\dbar \rightarrow W^+ \ s\sbar + ng$
and $u\dbar \rightarrow W^+ \ u\ubar + ng$.
The conclusions stated in the previous paragraph about the $Z$ 
boson were found to be equally applicable to the $W$ boson,
with the exception that the $W$ boson was found to play a larger role in evaluation speed.

\section{Subprocess Cross-Sections in Multi-Jet Production}\label{mlm}

\begin{figure}
\resizebox{0.5\textwidth}{!}{%
  \includegraphics{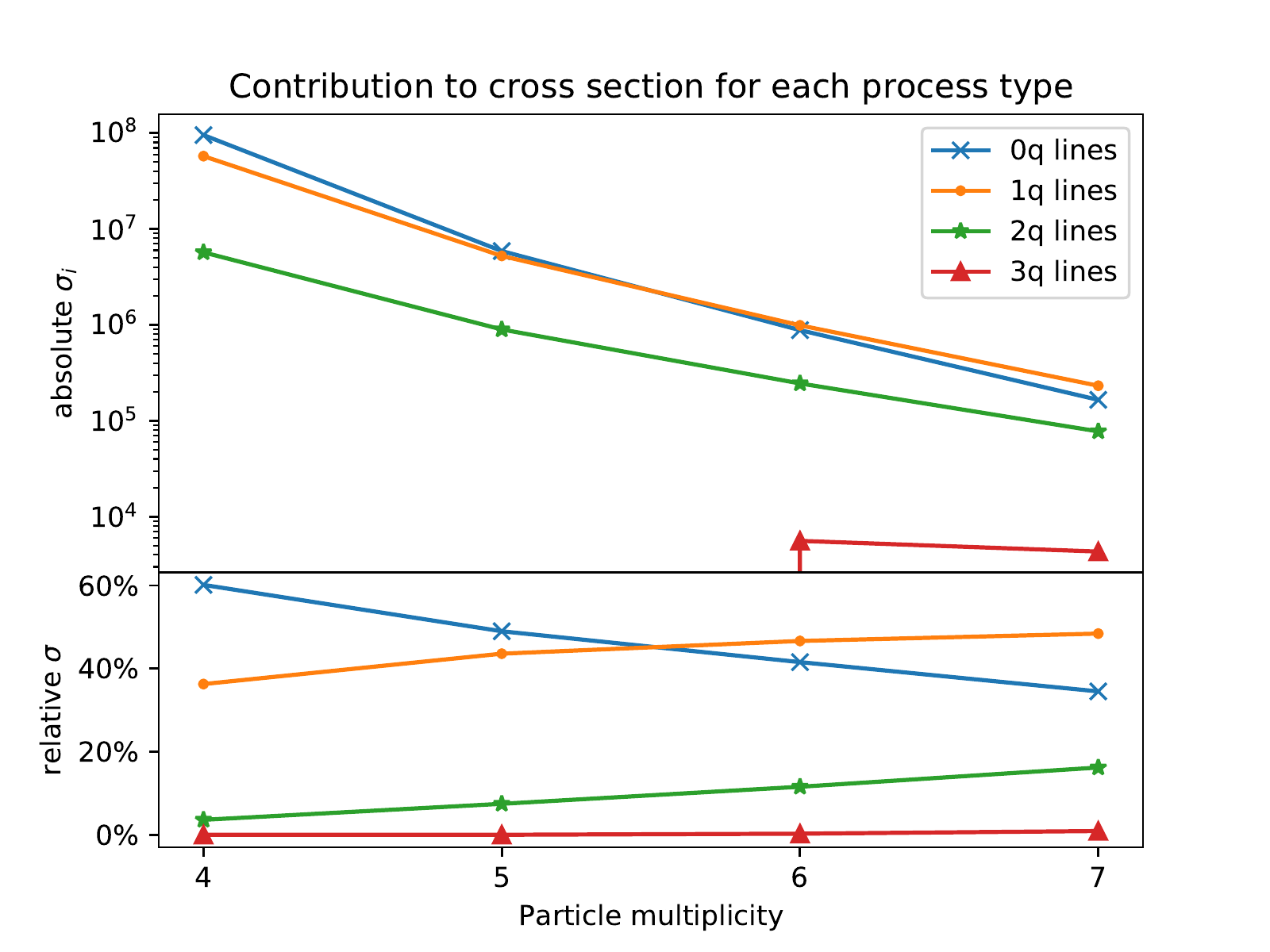}
}
\caption{MLM cross-section for each multiplicity and amplitude type without the inclusion of the Sudakov form-factor ($\textit{xqcut}=10\textrm{GeV})$.
% .\OM{Can we have the ration plot normalised to the total cross-section (and not the cross-section by multiplicity)?}
}
\label{fig:cross-section}       % Give a unique label
\end{figure}

In this appendix, we study the relative importance of various types of subprocesses 
classified by the number of quark lines present in the sub-process. 
Such information advises how critical it is to optimise the speed of the various contributions.

%(pure gluon, one quark line  for multi-jet
%Before comparing the speed of the various computations, 
%it is important to have an idea of the relative importance of each contribution to the total cross-section. 
In figure  \ref{fig:cross-section}, 
we present the tree-level cross-sections for multi-jet production, 
grouped via the number of quark lines present within the associated subprocess. 
The cross-sections are computed at partonic level as they would be within the MLM mode of \mgs{}.

The code used to perform his is the following:
\begin{lstlisting}[language=Bash]
 generate p p > 2j
 add process p p > 3j
 add process p p > 4j
 add process p p > 5j
 output
 launch
 shower=OFF
 set xqcut 10
\end{lstlisting}

As can be seen from the above set of commands, 
everything is default except for the value of xqcut,
and therefore the only additional cut is the maximum rapidity of the jet which is set at 5. 
Additionally, the PDF is NNPDF 2.3 (lhaid=247000) \cite{Ball:2013hta}.
We stress that within this procedure, no Sudakovs are included at parton-level.
Those factors are normally included after the running of the parton-shower by vetoing some of the generated events.
Therefore the reported cross-sections contain double-counting and should not be compared to experimental results. 
However, these values dictate how many events need to be generated within each category and therefore indicate the relative-importance of each category for a typical multi-jet calculation.

From figure \ref{fig:cross-section}, we can conclude that, like in any matched-merged computation, 
the cross-section is dominated by the lowest multiplicity,
which is fast and easy to compute.
However, the overall computation time is dominated by the highest multiplicity sample due to lower event generation efficiency and slower matrix-element evaluation.

 At high multiplicity, the full gluon amplitude is second to (but basically on par with) the single quark line (that includes 
 $gg\to q\bar q(n-4)g$, $q\bar q \to (n-2)g$ and  $q/\bar q g \to q/\bar q (n-3)g$). 
 Higher numbers of quark lines are suppressed at such multiplicities,
 with the three-quark line being completely negligible. 
 By extrapolating the plot for higher multiplicity, one can guess that cross sections with two quark lines will surpass the full gluon amplitude at either multiplicity 8 ($2\to6$ process) or 9.

\section{Modified Colour Expansion for Multiquark Amplitudes}
\label{sec:modLC multiquark}
Here we describe an attempt to modify the colour expansion in multiquark amplitudes,
for reasons outlined at the end of \secref{sec:1/nc expansion}.
It was found that this modified colour expansion does not overly help the accuracy or precision of the colour expansion,
but we leave it here for the interested reader.

% \commentAL{Keep this summary?}\OM{yes}
For two quark pairs, a strict colour expansion includes the $1/\Nc$ 
from the $\uone$ gluon in the expansion. 
This implies that only the second and third lines of \eqref{eq:4q amplitude fundamental}
are included at LC if the quark lines have different flavour. 
Therefore, unlike for single-quark or all-gluon amplitudes, we do not include all kinematic amplitudes at least once at LC.

We therefore propose a modified colour or `modN$^{k}$LC' expansion,
which does not count the $1/\Nc$ terms coming from the $\uone$ 
gluon in the expansion, 
but rather includes it in the definition of the kinematic amplitude.
The remaining rules of the expansion continue as before.

In other words, \eqref{eq:4q amplitude fundamental}
is changed to 
\begin{align}
\mathcal{\hat{M}}(q\bar{q}Q\bar{Q}+ng) &= 
\sum_{i=0,n}\sum_{P(1,\dots,i)}\sum_{P(i+1,\dots,n)} 
\left[\vphantom{\frac{1}{\Nc}}\right. \nonumber \\
&(t^1\dots t^i)_{q\bar{Q}}(t^{i+1}\dots t^n)_{Q\bar{q}}\nonumber \\
&\quad \times M(q,1,\dots,i,\bar{Q},Q,i+1,\dots,n,\bar{q}) \nonumber \\
&-(t^1\dots t^i)_{q\bar{q}}(t^{i+1}\dots t^n)_{Q\bar{Q}}\nonumber \\
&\left.\vphantom{\frac{1}{\Nc}}\quad\times M^*(q,1,\dots,i,\bar{q},Q,i+1,\dots,n,\bar{Q})\right]~,
\label{eq:4q amplitude modLC}
\end{align}
where we defined $M^* = \frac{1}{\Nc}M$.
In this way, the colour matrix and expansion ignores the colour suppression of the 
$\uone$ gluon and includes all kinematic amplitudes already at LC.

\begin{figure*}
\resizebox{1\textwidth}{!}{%
  \includegraphics{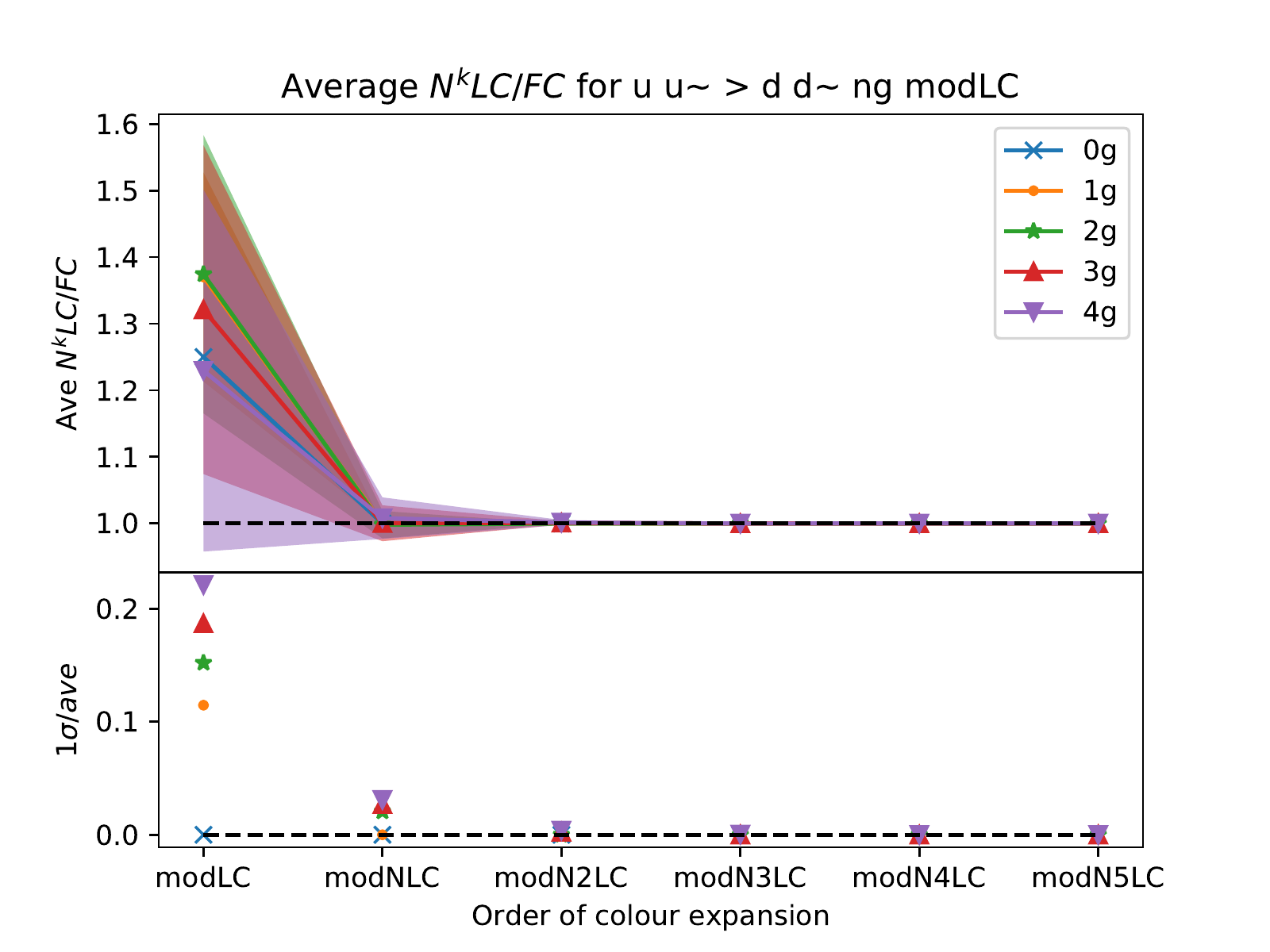}
  \hfil
  \includegraphics{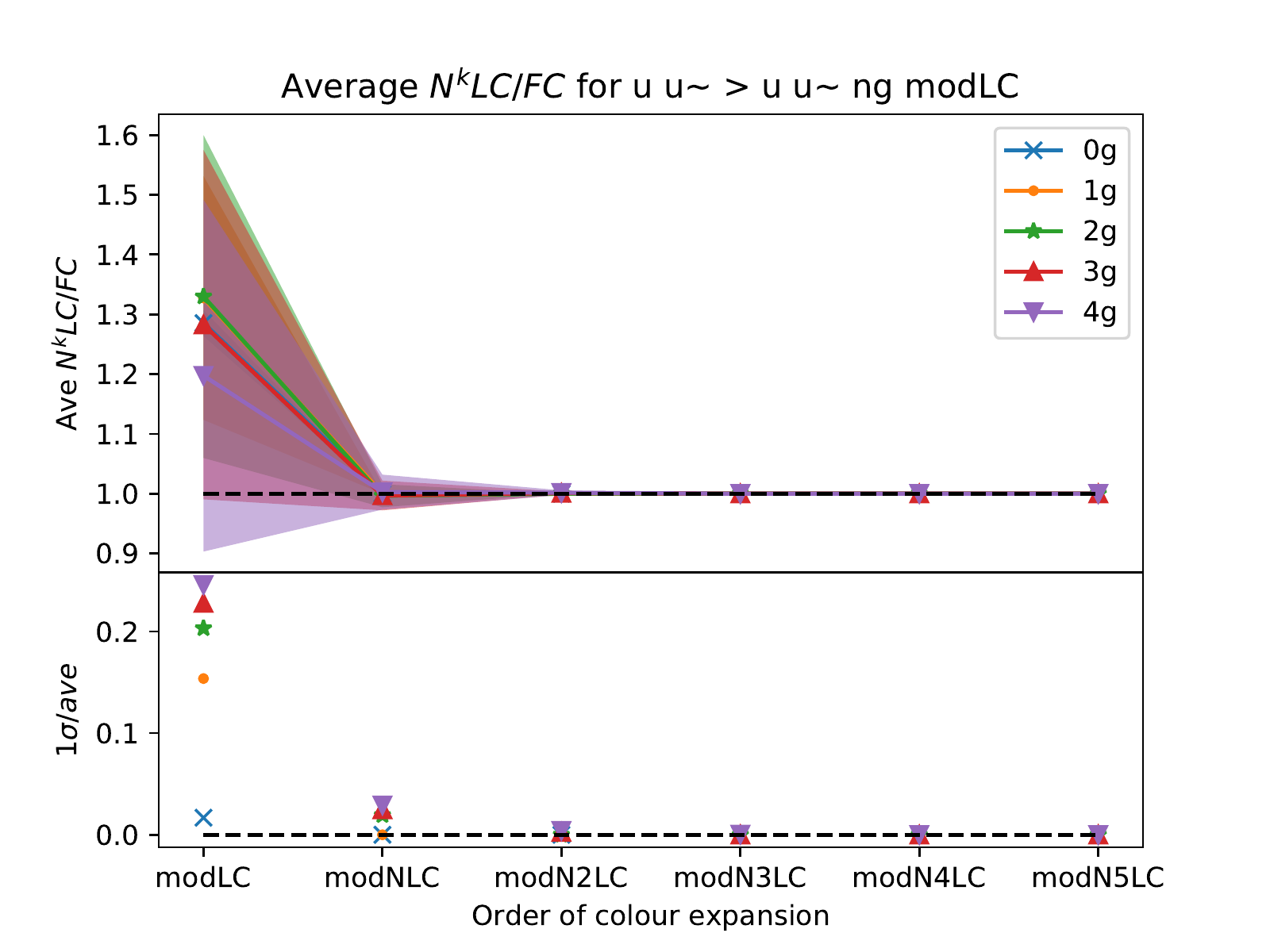}
}\caption{Same as \figref{fig:multiquark accuracy}
but at modified colour.}
\label{fig:multiquark modlc}
\end{figure*}

As we see by comparing \figref{fig:multiquark modlc} to \figref{fig:multiquark accuracy},
modified colour decreases the accuracy,
but has up to half the relative uncertainty at LC if the quarks have different flavours. 
Despite this positive effect, 
the modLC amplitudes are not precise enough for practical corrections.
In addition, the same-flavour precision actually gets worse using this colour expansion.

The speed of the modified expansion was found to be slower at each order but the same at the end of the expansion as expected.

% \begin{figure*}
% \resizebox{1\textwidth}{!}{%
% \begin{subfigure}{1\textwidth}
%   \includegraphics{figures/Accuracy/ave_std_NnLcDivFc_nospace_relerr_discrete_4qngDiffModLC.pdf}\hfill
%   \includegraphics{figures/speed/ave_speed_vs_ng_nospace_discrete_4quarkDiffModLC.pdf} \\
% \end{subfigure}
% \begin{subfigure}{1\textwidth}
%  \includegraphics{figures/Accuracy/ave_std_NnLcDivFc_nospace_relerr_discrete_4qngSameModLC.pdf}\hfill
%   \includegraphics{figures/speed/ave_speed_vs_ng_nospace_discrete_4quarkSameModLC.pdf}
% \end{subfigure}
% }\caption{Same as \figsrefa{fig:all-gluon accuracy}{fig:all-gluon speed} 
% but for $u\ubar \rightarrow d\dbar + ng$ (left) and $u\ubar \rightarrow u\ubar + ng$ (right)
% both at modified colour.}
% \label{fig:multiquark modlc}
% \end{figure*}

% % For one-column wide figures use
% \begin{figure}
% % Use the relevant command for your figure-insertion program
% % to insert the figure file.
% % For example, with the option graphics use
% \resizebox{0.75\textwidth}{!}{%
%   \includegraphics{leer.eps}
% }
% % If not, use
% %\vspace{5cm}       % Give the correct figure height in cm
% \caption{Please write your figure caption here}
% \label{fig:1}       % Give a unique label
% \end{figure}
% %
% % For two-column wide figures use
% \begin{figure*}
% % Use the relevant command for your figure-insertion program
% % to insert the figure file. See example above.
% % If not, use
% \vspace*{5cm}       % Give the correct figure height in cm
% \caption{Please write your figure caption here}
% \label{fig:2}       % Give a unique label
% \end{figure*}
% %
% % For tables use
% \begin{table}
% \caption{Please write your table caption here}
% \label{tab:1}       % Give a unique label
% % For LaTeX tables use
% \begin{tabular}{lll}
% \hline\noalign{\smallskip}
% first & second & third  \\
% \noalign{\smallskip}\hline\noalign{\smallskip}
% number & number & number \\
% number & number & number \\
% \noalign{\smallskip}\hline
% \end{tabular}
% % Or use
% \vspace*{5cm}  % with the correct table height
% \end{table}
%
% BibTeX users please use
\bibliographystyle{spphys}
\bibliography{mgbib}

\end{document}